\documentclass[sigconf]{acmart}
\usepackage{booktabs} 
\usepackage{framed}
\usepackage{amsthm}
\usepackage{amssymb}
\usepackage{bm}
\usepackage{bbm} 
\usepackage{lettrine}
\usepackage{framed}
\usepackage{xcolor}
\usepackage{algorithm}
\usepackage{algorithmic}
\usepackage{overpic}
\usepackage{framed}
\usepackage{dsfont}
\usepackage{graphicx}
\usepackage{caption}
\usepackage{subcaption}
\usepackage{algorithm}
\usepackage{algorithmic}
\usepackage{url}
\newcommand{\edit}[1]{{ #1}}

\AtBeginDocument{%
  \providecommand\BibTeX{{%
    \normalfont B\kern-0.5em{\scshape i\kern-0.25em b}\kern-0.8em\TeX}}}
\setcopyright{acmcopyright}
\copyrightyear{2020}
\acmYear{2020}
\acmDOI{10.1145/1122445.1122456}
\acmConference[Sigmetrics '20]{Sigmetrics '20}{June 08--12, 2020}{Boston, MA}
\acmBooktitle{Sigmetrics '20,
  June 08--12, 2020, Boston, MA}
\acmPrice{15.00}
\acmISBN{978-1-4503-9999-9/18/06}
\begin{document}
\title{Fundamental Limits of Online Network-Caching}
\author{Rajarshi Bhattacharjee}
\affiliation{\institution{Dept. of Electrical Engineering\\ IIT Madras}}
\email{brajarshi91@gmail.com}
\author{Subhankar Banerjee}
\affiliation{\institution{Dept. of Electrical Engineering\\ IIT Madras}}
\email{ee16s048@ee.iitm.ac.in}
\author{Abhishek Sinha}
\affiliation{\institution{Dept. of Electrical Engineering\\ IIT Madras}}
\email{abhishek.sinha@ee.iitm.ac.in}
%
%
%
%
%
%
\begin{abstract}
Optimal caching of files in a content distribution network (CDN) is a problem of fundamental and growing commercial interest. Although many different caching algorithms are in use today, the fundamental performance limits of network caching algorithms from an online learning point-of-view remain poorly understood to date. In this paper, we resolve this question in the following two settings: (1) a single user connected to a single cache, and (2) a set of users and a set of caches interconnected through a bipartite network. Recently, an online gradient-based \emph{coded} caching policy was shown to enjoy sub-linear regret. However, due to the lack of known regret lower bounds, the question of the optimality of the proposed policy was left open. In this paper, we settle this question by deriving tight non-asymptotic regret lower bounds in both of the above settings. In addition to that, we propose a new Follow-the-Perturbed-Leader-based \emph{uncoded} caching policy with near-optimal regret. Technically, the lower-bounds are obtained by relating the online caching problem to the classic probabilistic paradigm of \emph{balls-into-bins}. Our proofs make extensive use of a new result on the expected load in the most populated half of the bins, which might also be of independent interest. We evaluate the performance of the caching policies by experimenting with the popular \textsf{MovieLens} dataset and conclude the paper with design recommendations and a list of open problems. 
\end{abstract}

\ccsdesc[500]{Networks~Network performance analysis}
\ccsdesc[500]{Mathematics of computing~Probabilistic algorithms}

\keywords{network-caching, online algorithms, regret bounds, fundamental limits}

\maketitle
\section{Introduction and Related work}\label{intro}
\lettrine[]{\textbf{T}}{ }he classical caching problem, which seeks to make popular contents quickly accessible by prefetching them in low-latency storage, has been extensively studied in the literature.  The core idea of caching has been used in many diverse domains, including improving the CPU paging performance via $L_1/L_2$ caches \cite{silberschatz2006operating}, web-caching by Content Distribution Networks \cite{aggarwal1999caching, nygren2010akamai, amazon2015amazon}, and low-latency wireless video delivery through Femtocaching \cite{shanmugam2013femtocaching}. With the exponential growth of internet video traffic and the advent of new services consuming high bandwidth, such as augmented and virtual reality (AR/VR), the importance of caching for ensuring the quality of service (QoS) is on the rise \cite{chakareski2017vr}. Top CDN providers, such as Amazon AWS and Microsoft Azure, now offer caching as a service \cite{varia2014overview, chappell2010introducing}.    

Several caching algorithms have been proposed in the literature. The  \textsf{MIN} algorithm \cite{van2007short} is an optimal \emph{offline} caching policy which assumes that the entire file request sequence is known non-causally in advance. \textsf{MIN} is often used as a benchmark for comparing the performance of the online caching policies.  Among the online policies, the Least Recently Used policy (\textsf{LRU}), the Least Frequently Used policy (\textsf{LFU}) \cite{lee1999existence}, the \textsf{FIFO} policy \cite{dan1990approximate}, and the online coded caching policy \cite{pedarsani2016online} have been studied extensively. However, the performance guarantees available for most of the online caching policies are highly contingent upon some a priori assumptions on the generative model of the file request sequence \cite{breslau1999web, Zipf}. The paper \cite{jelenkovic2008characterizing} analyzed the performance of the \textsf{LRU} caching policy with an i.i.d.\ file request sequence, also known as the Independent Reference Model \cite{vanichpun2004output}. Under a Markovian assumption, the paper \cite{pedarsani2016online} shows that a coded caching policy out-performs the \textsf{LRU} policy. The paper \cite{flajolet1992birthday} develops a unified framework for analyzing a number of popular caching policies, again with a stationary file request model.  On a different line of work, the papers \cite{maddah2014fundamental, caching_rate1, caching_rate2, caching_rate3} derive information-theoretic lower bounds and efficient caching, computing, and coding schemes to facilitate bandwidth-efficient delivery of the cached files to the users. 

With frequent addition of new content to the library, mobility of the users, Femtocaching with small caches, and change in the popularity distribution with time, the assumption of stationary file popularity barely holds in practice \cite{traverso2015unravelling}. This prompts us to consider the problem of caching from an online learning point-of-view with no a priori statistical assumption on the file request sequence. Our work is inspired by the recent paper \cite{paschos2019learning}, which describes an online gradient-based coded caching policy (OGA), and proves a  sub-linear regret upper-bound for the same. Interestingly, they also show that popular uncoded caching policies, such as \textsf{LRU}, \textsf{LFU}, and \textsf{FIFO}, suffer from linear regrets in the worst case. In fact, no uncoded caching policy with a sub-linear regret is known previously in the literature. More seriously, no regret lower bound is known for the network-caching problem. 

In contrast to the multi-armed bandits setting \cite{DBLP:conf/colt/MagureanuCP14, combes2015learning, kveton2015tight, grunewalder2010regret, agrawal2013further}, relatively few results are known for the regret lower bounds for online convex optimization problems. Technically, the network-caching problem is an instance of an online convex optimization problem with a piecewise linear reward function and polytope constraints. The paper \cite{Abernethy08optimalstrategies} establishes a minimax regret lower bound for linear cost functions with hyper ball constraints. A regret lower bound for the  \emph{unconstrained} linear cost functions has been obtained in \cite{hazan2006efficient}. The papers \cite{hazan2007logarithmic, hazan2014beyond} prove logarithmic regret bounds for online stochastic strongly convex problems. \edit{However, to the best of our knowledge, with the exception of \cite{paschos2019learning}, the regret for a linear cost function with a simplex and several box constraints, which arise in the context of the single cache problem, has not been studied before. Moreover, the problem of lower bounding the regret for a piecewise linear cost function with polytope constraints, which arise in the context network-caching, is completely open.}

\edit{The above considerations inspire us to ask the following two questions in this paper:\\
\textbf{Question 1.} What is the fundamental performance limit of all online caching policies \emph{regardless} of their operational constraints or computational complexity?\\
\textbf{Question 2.} Can a simple, distributed network-caching strategy be designed which meets the above fundamental limit?

 In answering Question 1, we derive universal regret lower bounds that also apply to computationally intensive caching policies, which can completely change the profile of the cached contents at \emph{every} time slot. Surprisingly enough, we answer Question 2 in the affirmative. In particular, our matching upper and lower bounds reveal that a simple gradient-based incremental coded caching policy is regret-optimal. Moreover, we propose a new Follow-the-Perturbed-Leader-based uncoded caching policy that has near-optimal regret. Hence, one of the key take-away points from this paper is that there exist computationally cheap caching policies that perform excellently in an online setting even with adversarial request sequence. 
 }

\paragraph{Our contributions:}
In the process of answering the above questions, we make the following key  technical contributions:
\paragraph{\bf (1) Lower bound for a single cache:} In Theorem \eqref{reg_lb_gnl}, we prove a tight non-asymptotic regret lower bound for the single-cache problem. This result improves upon the previously known \emph{asymptotic} regret lower bound in \cite{paschos2019learning}, which can be arbitrarily loose for a sufficiently large library size. 
\paragraph{\bf (2) Lower bounds for caching networks:} In Theorems \eqref{elastic_th} and \eqref{inelastic_th}, we derive non-asymptotic sub-linear regret lower bounds for bipartite caching networks. We also show that the lower bounds are tight within constant factors. 
To the best of our knowledge, this is the first known regret lower bound for piecewise linear functions with polytope constraints. Hence, our results also contribute to the growing literature on online convex optimization.
\paragraph{\bf (3) Near-optimal uncoded caching policy:} Although the above lower bounds certify the optimality of the gradient-based coded caching policy of \cite{paschos2019learning}, it was an open problem whether one could achieve the optimal regret without coding. In Algorithm \ref{uncoded}, we propose a simple uncoded caching policy based on the Follow-the-Perturbed-Leader (\textsf{FTPL}) paradigm. Theorem \ref{uc_achievability} shows that the  \textsf{FTPL} policy has near-optimal regret. 
\paragraph{\bf (4) New proof techniques:} \edit{Technically, the regret lower bounds are established by relating the online caching problem to the classic probabilistic setup of \emph{balls-into-bins} via Lemma \ref{max_load}. In this Lemma, we derive a non-asymptotic lower bound to the expected total load in the most populated $n$ bins when $m$ balls are randomly thrown into $2n$ bins. For our lower bound proofs, we are particularly interested in the regime $m \gg n$. It is to be noted that classical results on randomized load balancing, such as \cite{mitzenmacher1996power, mitzenmacher2017probability, gonnet1981expected}, do not apply to this setting because they apply only to the regime where $m=O(n).$ It is to be noted that the paper \cite{ballsinbins} derives an asymptotic high-probability bound for the maximum load for various asymptotic regimes of $m$ and $n$. However, since we are primarily interested in the \emph{non-asymptotic expected} value of Max-Load, the results of \cite{ballsinbins} do not suffice for our purpose. Consequently, we tackle this problem from the first principles, culminating in Lemma \ref{max_load}.} To the best of our knowledge, this is the first paper where a connection between online learning and the framework of balls-into-bins has been explicitly brought out and exploited in proving regret lower bounds. 
\edit{\paragraph{\bf (5) Numerical experiments}	In Section \ref{experiments}, we compare the performance of different caching policies using the popular MovieLens 1 M dataset \cite{movieLens}. Our experiments reveal that the proposed \textsf{FTPL} policy beats other competitive caching policies in terms of long-term average regret. \\}

\section{Single Cache} \label{single_cache_section}
In this section, we begin our investigation with the single cache problem. We establish a key technical lemma on the 
\emph{balls-into-bins} problem, which is used in all of our lower bound proofs. Our analysis in this section improves upon the best-known regret lower bound for the single caches \cite{paschos2019learning} and paves the way for analyzing the bipartite caching problem in Section \ref{net_cache}. 
\subsection{System Model} \label{system_model}
\begin{figure}
\centering
\begin{overpic}[width=0.42\textwidth]{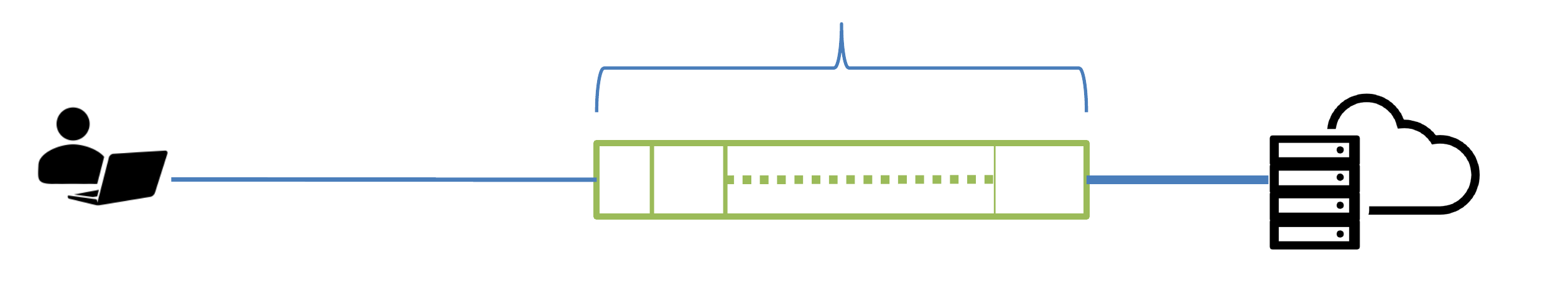}
\end{overpic}
\put(-130,40){\footnotesize{Cache capacity $=C$}}
\put(-35,-2){\footnotesize{Server}}
\put(-208, 0){\footnotesize{User}}
\caption{\small {Caching with a single cache}}
\label{single_cache_fig}
\end{figure}

In the classical caching problem with a single cache, there is a library of $N$ distinct files. A cache with a limited storage capacity can store at most $C$ files at any time slot. In practice, the cache capacity is significantly smaller compared to the entire library size (\emph{i.e.,} $C\ll N$). \edit{As an example, we may think of caching movie files in the Netflix data center, where the library size $N$ increases every day as new movies are released, but the physical cache-size in the Netflix data centers remains constant on the relevant time-scale.}  Time is slotted, and a user may request at most one file at a time. The file requests at time slot $t$ is represented by an $N$-dimensional binary vector $\bm{x}_t \in \{0,1\}^N$, where $x_{tf}=1$ if the $f$\textsuperscript{th} file is requested by the user at time $t$, and is zero otherwise (one-hot encoding). Following an online caching policy $\pi$, files are cached at every time slot \emph{before} the request for that slot arrives.  We do not make any statistical assumption on the file request sequence $\{\bm{x}_t\}_{t \geq 1}$. Thus, we may as well assume that the requests are made by an omniscient adversary who has complete knowledge of the cached contents and the caching policy in use.  See Figure \ref{single_cache_fig} for a schematic.
\paragraph{Coded Caching:} In this paper, we consider both coded and uncoded caching. In the classical uncoded caching, complete files are cached. On the other hand, in coded caching, the original files are first encoded using fountain codes (a class of rateless erasure codes), e.g., Raptor code \cite{shokrollahi2006raptor, luby2011raptorq}, and then some of the resulting coded symbols are cached. These codes have the property that an original file consisting of $k$ source symbols can be recovered (with high probability) by combining any subset of $k'$ coded symbols, where $k'$ needs to be only slightly larger than $k$. Hence, for decoding, it does not matter which encoding symbols are combined, as long as the decoder has access to sufficiently many encoded symbols. We will see that coding offers distinct advantages for network caching. These codes also admit highly efficient linear time encoding and decoding operations. The rateless codes are routinely used in P2P data streaming, large scale data centers, and CDNs \cite{wu2007rstream}. 
\paragraph{Caching configuration:}
The cache configuration at time $t$ is represented by an $N$-dimensional vector $\bm{y}_t^\pi \in [0,1]^N$, where $y_{t,f}^\pi$ denotes the fraction of the file $f$ cached at time $t$ under the policy $\pi$ \footnote{Whenever the caching policy $\pi$ is clear from the context, we will drop the superscript $\pi$ to simplify the notation.}. Naturally, in the uncoded case $y^{\pi}_{t,f} \in \{0,1\}, \forall f,t$. The set of all admissible caching configuration is denoted by $\mathcal{Y}$ where 
\begin{eqnarray} \label{admissible}
\mathcal{Y}= \bigg\{ \bm{y} \in [0,1]^N: \sum_{f=1}^N y_f \leq C \bigg\}, 
\end{eqnarray}
 where $C$ is the capacity of the cache. The caching decision $\bm{y}_t$ may be randomized and may depend on the file request sequence and caching decisions up to time $t-1$. Any requested file, not present in the cache, is routed to a central server and accrues zero reward.   

\subsection{Reward}
 A popular performance metric for any online caching policy is its average \emph{hit rate}, \emph{i.e.,} the average number of requested files already present in the cache so that the files can be quickly retrieved.  
 In this connection, let the user-generated file request sequence be denoted by $\{\bm{x}_t\}_{t\geq 1}$. The total reward up to time $T$ accrued by a caching policy, which responds to the file requests by setting the cache configuration to $\bm{y}_t$, at time $t$, is denoted by $Q\big(\{\bm{x}_t\}_{1}^{T}, \{\bm{y}_t\}_{1}^{T}\big)$.  We assume that the cumulative reward over a time horizon $T$ has an additive structure, which may be  obtained by summing up the rewards obtained at every slot up to time $T$, \emph{i.e.,}
\begin{eqnarray} \label{reward_def}
	Q\big(\{\bm{x}_t\}_{1}^{T}, \{\bm{y}_t\}_{1}^{T}\big) = \sum_{t=1}^{T} q(\bm{x}_t, \bm{y}_t).
\end{eqnarray}
The one-slot reward function $q(\cdot, \cdot)$ captures the reward obtained per slot. Intuitively, $q(\bm{x}_t,\bm{y}_t)$ denotes the extent of cache-hits for the request vector $\bm{x}_t$ against the cache configuration vector $\bm{y}_t$. The function $q(\cdot, \cdot)$ takes different functional forms depending on whether we consider (a) single cache, or (b) a network of caches. 
For the case of a single cache, we define the one-slot reward to be the amount of the requests successfully served by the cache, \emph{i.e.,}
\begin{eqnarray} \label{single_cache}
q(\bm{x}_t, \bm{y}_t) \equiv \bm{x}_t \cdot \bm{y}_t.\end{eqnarray}
A generalized definition of the above one-slot reward function applicable to caching networks will be given in Section \ref{net_cache}.
  
\subsection{Regret}
Since we do not make any assumption on the user-generated file request sequence $\{\bm{x}_t\}_{t\geq 1}$, it is futile to attempt to optimize the total reward (\emph{i.e.,} hit rate). This is because, at any slot, an omniscient adversary can always request a file which is not present in the cache, thus yielding a total of zero hits. To obtain a non-trivial performance measure, we cast the caching problem into the framework of online learning. This prompts us to compare the performance of any policy with the best offline \emph{stationary} optimal policy \cite{shalev2012online}. Let the vector $\bm{y}^*$ denote any fixed stationary cache configuration vector. The vector $\bm{y}^*$ may be selected offline after seeing the entire request sequence $\{\bm{x}_t\}_{t=1}^{T}$. Following the usual convention in the online learning literature, we define the regret $R^\pi_T(\{\bm{x}_t\}_{1}^{T})$ for a request sequence $\{\bm{x}_t\}_{1}^{T}$ to be the difference in the reward obtained by the best stationary caching configuration $\bm{y}^*$ and that of the online policy $\pi$. Mathematically\footnote{Recall that the cache configuration sequence $\{\bm{y}_t\}_{t\geq 1}$ is determined by the policy $\pi$.}, 
\begin{eqnarray} \label{regret_seq}
R^\pi_T(\{\bm{x}_t\}_{1}^{T}) := \sup_{\bm{y}^* \in \mathcal{Y}}\bigg( Q\big(\{\bm{x}_t\}_{1}^{T}, \{\bm{y}^*\}_{1}^{T}\big) - Q\big(\{\bm{x}_t\}_{1}^{T}, \{\bm{y}_t\}_{1}^{T}\big)\bigg).  
\end{eqnarray}
The regret $R^\pi_T$ of any caching policy $\pi$ up to time $T$ is defined to be the maximum regret over all admissible request sequences, \emph{i.e.,}
\begin{eqnarray} \label{regret_def1}
R^\pi_T := \sup_{\{\bm{x}_t\}_{1}^{T}}  R^\pi_T(\{\bm{x}_t\}_{1}^{T}).
\end{eqnarray}

\subsection{Regret lower bounds - preliminaries} \label{basic}
Using the simple observation that the maximum of a set of real numbers is at least equal to their average, for any joint probability distribution $P(\{\bm{X_t}\}_{t=1}^{T})$ on the file request sequence $\{\bm{X_t}\}_{t=1}^{T}$, the regret in Eqn.\ \eqref{regret_def1} may be lower bounded as follows:
\begin{eqnarray} \label{unif_lb}
	  R^\pi_T \geq \mathbb{E}_{\{\bm{X}_t\}_{1}^{T}} R^\pi_T(\{\bm{X}_t\}_{1}^{T}).
\end{eqnarray}
The joint distribution $P(\cdot)$ needs to be chosen carefully to ensure the tractability of evaluating the expectation in \eqref{unif_lb}, as well as the tightness of the resulting bound. The above technique is known as the \emph{probabilistic method} popularized by Erd\H{o}s \cite{alon2004probabilistic}.
In all of our proofs, we lower bound the quantity in Eqn.\ \eqref{regret_seq} by a suitable \emph{binary} cache configuration vector $\bm{y}^*$ (thus $\bm{y}^*$ corresponds to uncoded caching). As a consequence, all of our lower bounds remain valid in both uncoded and coded caching.

\subsection{Regret lower bound for unit-sized cache}
%

We first consider a single cache of unit capacity and derive a universal non-asymptotic regret bound. As we will see in the sequel, understanding this special case lays the foundation to our subsequent analysis of caching networks serving multiple users. 
\begin{framed}
\begin{theorem}[Lower bound for a single-cache of unit capacity] \label{reg_lb_spl}
The regret $R^\pi_T$ of any online caching policy, for a library-size $N=2$ and cache-capacity $C=1$, is lower bounded as: \[R^\pi_T \geq \sqrt\frac{T}{2\pi} - \frac{1}{2\sqrt{2\pi T}}, ~~\forall T\geq 1.\]
\end{theorem}
\end{framed}

\begin{proof}
Denote the cache configuration selected by the caching policy $\pi$ at slot $t$ by the vector $\bm{y}_t=\begin{pmatrix}\gamma_t && 1-\gamma_t\end{pmatrix}'$, where $\gamma_t$ denotes the fraction of $\textsf{File}_1$ cached by the policy $\pi$ ($0 \leq \gamma_t \leq 1$). Recall the definition of regret $R_T$ in this context:
\begin{eqnarray} \label{regret_def}
R_T^\pi=\sup_{\{\bm{x}_t\}_{1}^{T}}\sup_{\bm{y}^* \in \mathcal{Y}} \bigg({\bm{y^*}}\cdot\sum_{t=1}^{T} \bm{x}_t - \sum_{t=1}^{T} \bm{y}_t \cdot \bm{x}_t\bigg),
\end{eqnarray}
where the set of all admissible configurations $\mathcal{Y}$ is given by Eqn.\ \eqref{admissible}. Denote the file request vector at slot $t$ by $\bm{x}_t=\begin{pmatrix} w_t && 1-w_t \end{pmatrix}'$, where $w_t \in \{0,1 \}$. It can be easily seen that for a given file request sequence $\{\bm{x}_t\}_{1}^{T}$, an optimal choice of the fixed cache configuration vector $\bm{y}^*$ in Eqn. \eqref{regret_def} is given as follows: 
\begin{eqnarray*}
\bm{y}^* = \begin{cases}
 \begin{pmatrix}
 1 && 0 	
 \end{pmatrix}',
 ~\textrm{if} ~ \sum_{t=1}^T w_t \geq T/2,\\
 \begin{pmatrix}
 0 && 1 
 \end{pmatrix}',
	~\textrm{if} ~ \sum_t w_{t=1}^T < T/2.
\end{cases}
\end{eqnarray*}
To prove a universal regret lower bound, we need to show the existence of an adversarial file request sequence $\{\bm{x}_t\}_{1}^{T}$ under which the caching policy $\pi$ performs poorly. Towards this, let $\{W_t\}_{t \geq 1}$ be a sequence of i.i.d.\ uniform Bernoulli random variables such that $\mathbb{P}(W_t=0)=\mathbb{P}(W_t=1)=\frac{1}{2}$. Construct a random file request sequence $\bm{X}_t=\begin{pmatrix} W_t && 1-W_t \end{pmatrix}'$. 
The regret incurred for the sequence $\{\bm{X}_t\}_{1}^{T}$ may be obtained from Eqn.\ \eqref{regret_def} as: 
\begin{eqnarray*}
&&R_T^\pi(\{\bm{X}_t\}_t)\\
 &=& \max \bigg\{\sum_{t=1}^T W_t, T - \sum_{t=1}^T W_t\bigg\} - \sum_{t=1}^T \big( \gamma_t W_t + (1-\gamma_t)(1-W_t)\big)\\
  &=& \max \big\{Z, T- Z\big\} - \sum_{t=1}^T \big( 2\gamma_t W_t + 1 - \gamma_t -W_t\big),
\end{eqnarray*}
where the r.v. $Z \equiv \sum_{t=1}^{T} W_t$, being the summation of $T$ i.i.d.\ uniform Bernoulli variables, is Binomially distributed with parameter $(T, 1/2)$. Using linearity of expectation, we can write 
\begin{eqnarray}\label{md_eqn1}
\mathbb{E}_{\{\bm{X}_t\}_{1}^{T}} \big(R^\pi_T(\{\bm{X}_t)\}_{1}^{T}\big) = \mathbb{E}\big(\max \big\{Z, T- Z\big\}\big) - T/2,	
\end{eqnarray}
where we have used the fact that $\mathbb{E}(W_t)=1/2$ and the caching decision $\gamma_t$ is independent of the incoming request $\bm{X}_t$. Observe that, we can write  
\begin{eqnarray} \label{md_eqn2}
\max \big\{Z, T- Z\big\} &=& \frac{T}{2} + |Z-T/2|.  
\end{eqnarray}
Thus, combining Eqns.\ \eqref{md_eqn1} and \eqref{md_eqn2}, we have 
\begin{eqnarray}\label{lb_1}
	\mathbb{E}_{\{\bm{X}_t\}_{1}^{T}} \big(R_T^\pi(\{\bm{X}_t)\}_{1}^{T}\big) = \mathbb{E}\big|Z-T/2\big|.
\end{eqnarray}
The mean absolute deviation for a symmetric binomial random variable may be computed in closed form by using De Moivre's formula (\cite{berend2013sharp}, Eqn.\ (1)) as follows:
\begin{eqnarray} \label{DeMoivre}
\mathbb{E}\big|Z-\frac{T}{2}\big| = \frac{1}{2^{T}}\bigg(\lfloor\frac{T}{2}\rfloor +1 \bigg)\binom{T}{\lfloor{\frac{T}{2}}\rfloor + 1}.
\end{eqnarray}
Eqn.\ \eqref{DeMoivre}, in combination with non-asymptotic form of Stirling's formula \cite{robbins1955remark}, yields the following \emph{non-asymptotic} lower bound
\begin{eqnarray}\label{MD_bound}
 \mathbb{E}\big|Z-\frac{T}{2}\big| \geq \sqrt{\frac{T}{2\pi}}- \frac{1}{2\sqrt{2\pi T}}, ~~ \forall T\geq 1.
 \end{eqnarray}
For details of the above calculations, please refer to Appendix \ref{MD_bound_proof}. Equation \eqref{MD_bound}, coupled with Eqn.\ \eqref{lb_1}, shows the existence of a file request sequence $\{\bm{x}_t\}_{1}^{T}$ such that \[R_T^\pi(\{\bm{x}_t\}_{1}^{T}) \geq \sqrt{\frac{T}{2\pi}}- \frac{1}{2\sqrt{2 \pi T}}, ~~ \forall T\geq 1.\]  
\end{proof}
\textbf{Remarks:} It can be seen that the above proof and the lower bound in Theorem \ref{reg_lb_spl} continue to hold even if we let the library size to be $N\geq 2$. This observation follows by constructing a randomized file request sequence where the first two files are requested with probability $\frac{1}{2}$ each and other files are requested with zero probability.


\subsection{Lower bound for caches of arbitrary size} \label{arbit}
 We now extend the previous result to caches with arbitrary size $C \geq 1$. This extension is non-trivial. We will see in the sequel that our analysis naturally leads us to investigate a random variable arising in connection with the classic probabilistic framework of \emph{balls-into-bins}, where a number of balls are thrown uniformly and independently at random to some bins \cite{ballsinbins}. The following lemma, which might be of independent interest, gives a non-asymptotic lower bound to the total number of balls in the most populated half of the bins.  
\begin{framed}
\begin{lemma}[Total occupancy in the most popular half]\label{max_load}
	Suppose that $T$ balls are thrown independently and uniformly at random into $2C$ bins. Let the random variable $M_C(T)$ denote the  number of balls in the most populated $C$ bins. Then  
	\begin{eqnarray*}
	\hspace{-5pt}\mathbb{E}(M_C(T)) \geq \frac{T}{2} + \sqrt{\frac{CT}{2\pi}} - \frac{(\sqrt{2}+1)C^{3/2}}{2\sqrt{2 \pi T}}- \sqrt{\frac{2}{\pi}}\frac{C^2}{T}.\end{eqnarray*}
\end{lemma}
\end{framed}
\paragraph{Proof outline:} The proof proceeds by pairing up the bins to form \emph{super bins} (see Fig. \ref{superbins_fig}), and then selects the most-occupied bin in each super bin to obtain a lower bound on $M_C(T)$. Finally, we conclude the proof by appealing to the mean-deviation bound in Eqn.\ \eqref{MD_bound}.  Please refer to Section \ref{max_load_proof} for the complete proof of Lemma \ref{max_load}. 

To improve readability,  in the rest of the paper we will rephrase the above bound as  \[\mathbb{E}\big(M_C(T)\big) \geq \frac{T}{2} + \sqrt{\frac{CT}{2\pi}} - \Theta(\frac{1}{\sqrt{T}}),\]
with the understanding that an explicit form of the lower order terms may be obtained by using Lemma \ref{max_load}, if required. 
%
The following Theorem is the main result of this section. 
\begin{framed}
\begin{theorem}[Lower bound for a single cache of arbitrary capacity] \label{reg_lb_gnl}
The regret $R_T^\pi$ of any online caching policy $\pi$, for a library size $N$ and cache capacity $C$ with $N\geq 2C$, is lower bounded as 
\begin{eqnarray} \label{regret_bound}
R_T^\pi \geq \sqrt\frac{CT}{2\pi} - \Theta(\frac{1}{\sqrt{T}}), ~~\forall ~T\geq 1.
\end{eqnarray}
\end{theorem}
\end{framed}
\paragraph{Proof outline:} The proof proceeds along the lines of Theorem \ref{reg_lb_spl}, where the $t$\textsuperscript{th} file requested $\bm{X}_t$ is chosen independently and uniformly at random from the \emph{first} $2C$ files from the library. Then, we show that the reward accrued by the best fixed cache configuration $\bm{y}^*$ corresponds (in distribution) to the total number of balls in the most populated half of the bins. We then conclude the proof by appealing to Lemma \ref{max_load}. Please refer to Section \ref{reg_lb_gnl_proof} for the complete proof of Theorem \ref{reg_lb_gnl}.
\paragraph{Comparison with Theorem 1 of \cite{paschos2019learning}:} In the single cache setting, the paper \cite{paschos2019learning} establishes a rather loose \emph{asymptotic} regret lower bound of $\sqrt{\frac{C}{N\pi}} \sqrt{CT}$, which could be arbitrarily smaller than the lower bound given in Eqn.\ \eqref{regret_bound} when the library size $N$ is sufficiently large.  Theorem \ref{reg_lb_gnl} improves the result in \cite{paschos2019learning} in two ways. First, it proves a regret lower bound that is \emph{independent} of $N$. 
As a consequence, we will soon see that it implies that the gradient-based coded policy of \cite{paschos2019learning} is regret-optimal up to a constant factor. Theorem \ref{reg_lb_gnl} also implies near optimality of the uncoded \textsf{FTPL} policy described in the next section. 
Second, unlike the regret lower bound in \cite{paschos2019learning}, the bound in Eqn.\ \eqref{reg_lb_gnl} is \emph{non-asymptotic}, thus giving a valid lower bound for any $T \geq 1$.


\subsection{Achievability} \label{uc}
We note that many popular classical uncoded caching policies, such as \textsf{LRU}, \textsf{LFU}, and \textsf{FIFO}, have linear regrets (Proposition 1 of \cite{paschos2019learning}). This can be simply understood from the following example: consider the single cache setting with $N=2, C=1$ and an alternating file request sequence $\{\bm{x}_t\}_{t \geq 1}=\{1,2,1,2,1,2,\ldots \}$. Since any missed content is always immediately loaded to the cache, each of the above policies gets zero cumulative hits. On the other hand, caching either one of the files forever achieves a total of $\frac{T}{2}$ cumulative hits for a horizon of length $T$. Thus, all of the above policies have $\Omega(T)$ regret. This is surprising as the \textsf{LRU} and \textsf{FIFO} policies are known to have a finite competitive ratio \cite{albers}. To the best of our knowledge, no uncoded caching policy with sub-linear regret is known in the literature.
\paragraph{Achievability with uncoded caching:}
Making use of the theory of Online Structured Learning \cite{cohen2015following}, we now propose a simple Follow the Perturbed Leader (\textsf{FTPL})-based \emph{uncoded} caching policy, which achieves $O(\sqrt{T})$ expected regret against an oblivious adversary. The \textsf{FTPL} policy maintains a cumulative running count of the number of times a file was requested so far. This count is then perturbed by adding i.i.d. Gaussian noise of zero mean and an appropriate variance to each of the count values. Finally, at each time slot, the top $C$ files with the highest perturbed count are loaded to the cache. The \textsf{FTPL} policy is formally described below in Algorithm \ref{uncoded}. 

\begin{algorithm}
\caption{\textsf{FTPL} policy for {\sc Uncoded Caching}}
\label{uncoded}
\begin{algorithmic}[1]
\STATE $\textbf{\textsf{count}}\gets \bm{0}, \eta \gets \frac{1}{(4 \pi \log N)^{1/4}}\sqrt{\frac{T}{C}}$
\FOR {$t=1$ to $T$}
\STATE $\textbf{\textsf{count}} \gets \textbf{\textsf{count}} + \bm{x}_t$
\STATE Sample $\bm{\gamma}_t \sim \mathcal{N}(0,\bm{I}_{N \times N})$ 
\STATE $\textbf{\textsf{perturbed count}}\gets \textbf{\textsf{count}}+ \eta\bm{\gamma}_t$
\STATE Sort $\textbf{\textsf{perturbed count}}$ in decreasing order and load the top $C$ files in the cache.
  \ENDFOR
\end{algorithmic}
\end{algorithm} 
We prove the following achievability bound for the \textsf{FTPL} policy.

\begin{framed}
\begin{theorem}[Achievability with uncoded caching]\label{uc_achievability}
	In the single cache setting, the \textsf{FTPL} uncoded caching policy achieves the following upper bound for expected regret (the expectation is taken over the randomness of the algorithm)
	\[\mathbb{E}_{\{\bm{\gamma}_t\}_{t \geq 1}}\big(R_T^{\textsf{FTPL}}) \leq  1.51(\log N)^{1/4}\sqrt{CT}.\]
\end{theorem}
\end{framed}

The proof of Theorem \ref{uc_achievability} follows a similar line of arguments as the proof of Theorem 1 of \cite{cohen2015following}. However, in this paper, we tighten the regret upper bound of \cite{cohen2015following} further by a factor of $O(\sqrt{C})$. This improvement follows by taking into account the constraint that only one file is requested per slot and any feasible cache configuration respects a natural box constraint. This tightening is essential in order to match the \textsf{FTPL} upper bound with the lower bound given in Theorem \ref{reg_lb_gnl}. The details of the proof are given in Section \ref{uc_achievability_proof}.\\
The lower bound of Theorem \ref{reg_lb_gnl} shows that \textsf{FTPL} is regret-optimal up to poly-logarithmic factors in the library size $N$. In the following, we show that this extra poly-log factor may be removed if we allow coded caching.

\paragraph{Achievability with coded caching:} In \cite{paschos2019learning}, Corollary 2, the authors showed that an Online Gradient Ascent (OGA)-based single-server caching policy, serving one file-request per slot, achieves a regret of value at most $\sqrt{2CT}$. To avoid repetition, a description of the general version of the OGA policy in the context of  network caching will be given in Section \ref{achievability}, which subsumes the single cache case. 

\section{Caching in a Content Distribution Network} \label{net_cache}
\edit{We now begin investigating the problem of optimal caching in a Content Distribution Network (CDN). In this problem, there is a set of geographically distributed users who periodically request files to a content provider (\emph{e.g.,} Netflix). The content provider maintains a global network of data centers, each caching some files up to its capacity. A user's file-request may be served by its neighboring data centers if the requested file exists in any of the neighboring caches. This client-server architecture gives rise to a bipartite content distribution network with the set of users and the set of caches constituting its two parts \cite{shanmugam2013femtocaching}. For a detailed case-study on the above caching architecture, including the global distribution of the data centers and performance measurement in the context of Netflix CDN, please refer to \cite{bottger2018open}. 
In the following Section, we show how the tools and techniques, developed in the previous section for a single cache, may be generalized to address this more challenging  problem.}
\subsection{System Model}
\begin{figure}
\centering
\begin{overpic}[width=0.42\textwidth]{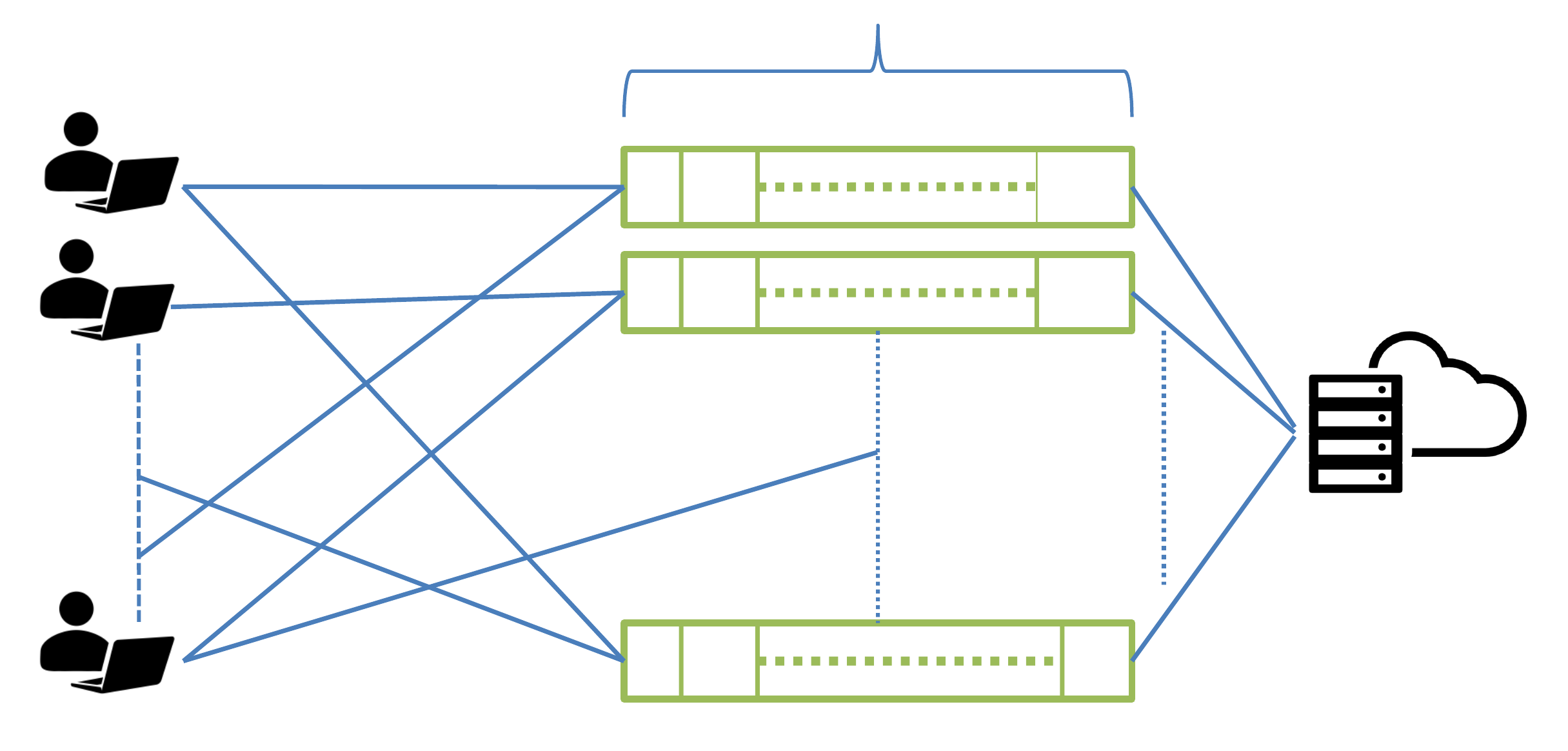}
\end{overpic}
\put(-125,100){\footnotesize{Cache capacity $=C$}}
\put(-30,25){\footnotesize{Server}}
\put(-210, -10){\footnotesize{$|\mathcal{I}|$ users}}
\put(-105, -10){\footnotesize{$|\mathcal{J}|$ caches}}
\put(-187, 85){\footnotesize{Connections $E$}}
\put(-180,-4){\footnotesize{right degree $=d$}}
\caption{\small {Schematic of a Bipartite Caching Network}}
\label{net_cache_fig}
\end{figure}
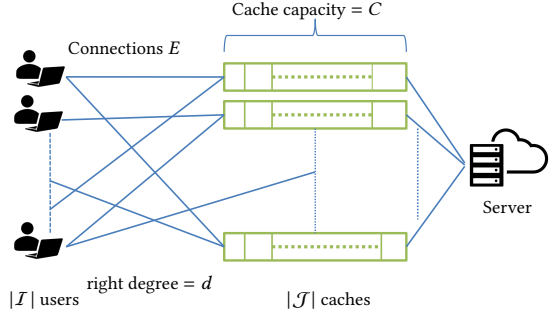

We now formalize the above system model for caching in a CDN. A set of users $\mathcal{I}=\{1,2,\ldots, I\}$ is connected to a set of caches  $\mathcal{J}=\{1,2, \ldots, J\}$ in the form of a bipartite network. 
For simplicity, we assume that the caches are homogeneous in the sense that each cache has the same storage capacity $C$. As before, the library size (\emph{i.e.,} the number of all possible files) $N$ is assumed to be sufficiently large. The connection between the users and the caches is represented by the bipartite graph $(\mathcal{I}, \mathcal{J}, E)$. The set of caches connected to a user $i \in \mathcal{I}$ is denoted by $\partial^+(i) \equiv \{j \in \mathcal{J}: (i,j) \in E\}$. Similarly, the set of users connected to a cache $j \in \mathcal{J}$ is denoted by $\partial^-(j) \equiv \{i \in \mathcal{I}: (i,j) \in E \}$. The \emph{in-degree} of the cache $j$ is defined as $d_j \equiv |\partial^{-}(j)|, j \in \mathcal{J}$. For the sake of simplicity, we assume the network to be right $d$-regular, \emph{i.e.,} $d_j=d, \forall j \in \mathcal{J}$. See Figure \ref{net_cache_fig} for a schematic.\\
Each user requests one file per time slot. Each file-request may be served by any (one or more) neighbouring caches. As before, the file request generated by a user $i$ at time $t$ is one-hot encoded by an $N$-dimensional binary vector $\bm{x}^i_t$, with the interpretation that $x^i_{tf}=1$ if and only if the $f$\textsuperscript{th} file is requested by the user $i$ at time $t$. The cache configuration of the $j$\textsuperscript{th} cache at time $t$ is represented by the $N$-dimensional vector $\bm{y}_t^j$, with each component denoting the fraction of the corresponding coded file cached. The cache configuration $\bm{y}_t$ must always satisfy the cache-capacity constraints. Thus, the set of all feasible cache configuration $\bm{Y}_{\mathcal{J}}$ is given by:
\begin{eqnarray} \label{feasibility}
 \bm{Y}_{\mathcal{J}}= \bigg\{ \big(\bm{y}^j, j\in \mathcal{J}\big): \sum_{f=1}^{N}y^j_f \leq C, \forall j \in \mathcal{J}, \bm{0}\leq \bm{y} \leq \bm{1}\bigg\}.
\end{eqnarray}
As before, for uncoded caching, we have $y^j_{ft}\in \{0,1\},~~\forall j,f,t.$
 

\subsection{Reward and Regret}
For content distribution networks, it will be useful to distinguish between \emph{elastic} and  \emph{inelastic} contents, as defined below. Making this distinction is essential due to the possibility that, in a caching network, the same content may be cached and retrieved from more than one cache at a time. Recall that, for rate less codes, it is only the \emph{total amount} of received encoded symbols that determine the decoding quality. 
\paragraph{Elastic contents:} We call a content to be \emph{elastic} if receiving multiple layers (\emph{i.e.,} resolutions) of the same content improves its overall utility for the users. Examples of elastic contents include multi-bitrate video files for adaptive streaming \cite{gu2013multiple}, \cite{adaptive_video}, multi-resolution HD videos \cite{holcomb2008multi}, and erasure-coded files in fault-tolerant distributed file system, such as Hadoop \cite{dimakis2010network}, \cite{shvachko2010hadoop}. In this setting, an incoming file request $\bm{x}_t^i$ from the $i$\textsuperscript{th} user can be satisfied by fetching and combining parts of the cached layers of contents from different neighboring caches $j \in \partial^+(i)$. Accordingly, for elastic contents, we define the one-slot reward to be the aggregate of the cache-hits at that slot, \emph{i.e.,}
\begin{eqnarray} \label{elastic_content}
	q_{\textsf{elastic}}(\bm{x}_t, \bm{y}_t) \equiv \sum_{i \in \mathcal{I}} {\bm{x}_t^i}\cdot \bigg(\sum_{j \in \partial^+(i)} \bm{y}_t^j\bigg). 
\end{eqnarray}
Hence, a user's utility increases linearly as she receives more layers of the requested content from different neighboring caches. 
\paragraph{Inelastic contents:} We call a content to be \emph{inelastic} if the content has only a single layer (resolution) and a user is fully satisfied if she is able to retrieve the original content. Popular examples of inelastic contents include traditional webpages, databases, documents, and single resolution images. Similar to the elastic case, if the files are encoded  by rate less MDS codes, a request from a user may be satisfied by combining different fractional parts of the content (from different neighboring caches),  with the fractions adding up to unity. Consequently, we define the one-slot reward for inelastic content to be 
\begin{eqnarray} \label{inelastic_content}
	q_{\textsf{inelastic}}(\bm{x}_t, \bm{y}_t) \equiv  \sum_{i \in \mathcal{I}}  {\bm{x}_t^i} \cdot \min \bigg\{\bm 1,\big(\sum_{j \in \partial^+(i)} \bm{y}_t^j\big)\bigg\},
\end{eqnarray}
where $\bm 1$ is the all-one vector, and the $\min (\cdot, \cdot)$ operator outputs a vector whose components are the pointwise minimum of the corresponding components of the input argument vectors.  
In comparison with the reward definition in Eqn.\ \eqref{elastic_content}, the $\min(\bm 1, \cdot)$ operator in Eqn.\ \eqref{inelastic_content} takes into account the fact that receiving multiple fractions of an inelastic content, summing up to more than one, does not add additional utility. Hence, unlike the elastic contents, inelastic contents have bounded rewards. We note that the reward in Eqn.\ \eqref{inelastic_content} coincides with the utility definition in Eqn.\ (13) of \cite{paschos2019learning}.   
The regret of any caching policy is defined exactly in the same way as in the single cache case via Eqns.\ \eqref{regret_seq} and \eqref{regret_def1}. 
\subsection{Achievability for Caching Networks} \label{achievability}
 
 In the following, we describe a simple and distributed gradient-based \emph{coded} caching policy that achieves $O(\sqrt{T})$ regret in both elastic and inelastic settings. We then propose an extension of the Follow the Perturbed Leader- based \emph{uncoded} caching policy given in Algorithm \ref{uncoded}, which also achieves near-optimal regret in the elastic setting.  
\paragraph{Achievability with coded caching \cite{paschos2019learning}:} Let $q(\bm{x}, \bm{y})$ be a generic one-slot reward function, which is concave in the cache-configuration vector $\bm{y}$ (\emph{e.g.,} $q(\cdot, \cdot)$ could be chosen to be either $q_{\textsf{elastic}}(\cdot, \cdot)$ or $q_{\textsf{inelastic}}(\cdot, \cdot)$). 
Let $\bm{g}_t$ be a supergradient of  $q(\bm{x}_t, \bm{y})$ at $\bm{y}=\bm{y}_t$. The paper \cite{paschos2019learning} describes the following Online Gradient Ascent (OGA)-based caching policy: starting from any initial feasible configuration $\bm{y}_0 \in Y_{\mathcal J}$, iterate as follows:
\begin{eqnarray} \label{bsa}
\bm{y}_{t+1}= \Pi_{Y_\mathcal{J}}\big( \bm{y}_t + \eta \bm{g}_t\big),	
\end{eqnarray}
where $\bm{Y}_\mathcal{J}$ is the set of all feasible cache configurations given in Eqn.\ \eqref{feasibility}, 
$\bm{\Pi}_{\bm{Y}_\mathcal{J}}(\cdot)$ is the Euclidean projection operator on the set $\bm{Y}_\mathcal{J}$, and $\eta >0$ is an appropriate step-size parameter. For the single user- single cache setting of Section \ref{single_cache_section}, we simply set $|\mathcal{I}|=|\mathcal{J}|=1.$
\paragraph{Distributed implementation:} The OGA-based caching policy can be implemented at each cache in a distributed fashion with locally available information only. This can be seen from the following two observations: 
\begin{enumerate}
\item A separable supergradient $\bm{g}_t$ for both the objective functions $q_{\textrm{elastic}}(\cdot, \cdot)$ and $q_{\textrm{inelastic}}(\cdot, \cdot)$ may be obtained, such that the gradient ascent steps in Eqn.\ \eqref{bsa} can be carried out locally. For an expression of such a supergradient, see Eqn.\ \eqref{sup_grad_expr}. 
\item Since the cache capacity constraints are separable for different caches, the projection operation $\bm{\Pi}_{\bm{Y}_\mathcal{J}}(\cdot)$ may be  carried out separately for each cache as
$\bm{\Pi}_{\bm{Y}_\mathcal{J}}(\bm{y}) = \{ \Pi_{\mathcal{Y}}(\bm{y}_j), ~~j \in \mathcal{J} \},$
where $\mathcal{Y}$ is the single cache constraint set given by Eqn.\ \eqref{admissible}. 
\end{enumerate}

We have the following achievability result for OGA:
\begin{framed}
\begin{theorem}[Achievability with coded caching \cite{paschos2019learning}]\label{achievability}
	For both elastic and inelastic contents, the OGA-based caching policy \eqref{bsa} with step size $\eta = \frac{\sqrt{2C}}{d\sqrt{T}} $, achieves the following upper-bound on regret for a right $d$-regular bipartite network:
	\[ R_T^{\textsf{OGA}} \leq d|\mathcal{J}|\sqrt{2CT}. \]
\end{theorem}
\end{framed}
\paragraph{Proof outline:} In this proof, we appeal to Theorem 2 of \cite{paschos2019learning},  which gives a generic regret upper bound for the OGA-based caching policy. We conclude the proof of Theorem \ref{achievability} by computing the diameter of the feasible set $Y_\mathcal{J}$ subject to the cache-capacity constraints. Note that we can not directly use the regret upper-bound given in Theorem $3$ of \cite{paschos2019learning} because there the authors make an assumption that only one user out of $|\mathcal{I}|$ users may request for contents at a slot. In our model, there is no such restriction so that all $|\mathcal{I}|$ users may simultaneously request for contents at a slot. Please refer to Appendix \ref{achievability_proof} for a proof of Theorem \ref{achievability}. 
\paragraph{Achievability with uncoded caching:} For uncoded caching in a bipartite network, we propose a simple extension of the \textsf{FTPL} policy given in Algorithm \ref{uncoded} for a single cache. In this extension, each of the $|\mathcal{J}|$ caches \emph{independently} implements the \textsf{FTPL} policy irrespective of whether the content is elastic or inelastic. We have the following achievability result for elastic contents:

  \begin{framed}
\begin{theorem}[Achievability with uncoded caching for elastic contents]\label{elastic_uncoded}
	For elastic contents, the \textsf{FTPL} caching policy with the noise parameter $\eta = \frac{d}{(4 \pi \log N)^{1/4}}\sqrt{\frac{T}{C}}$ yields the following upper-bound on expected regret for a right $d$-regular bipartite network:
	\[ \mathbb{E}_{\{\bm{\gamma}_t\}_t}(R_T^{\textsf{FTPL}}) \leq  1.51(\log N)^{1/4}d|\mathcal{J}|\sqrt{CT}. \]
	\end{theorem}
	\end{framed}
See Appendix \ref{elastic_uncoded_pf} for a proof of Theorem \ref{elastic_uncoded}. 
\subsection{Converse for Caching Networks} \label{converse}
The question of regret-optimality of the OGA policy \eqref{bsa} for caching networks was left open in \cite{paschos2019learning} due to lack of known lower bounds. In the following, we prove tight universal lower bounds for the regret, which applies to both coded and uncoded caching. 

\begin{framed}
\begin{theorem}[Lower bound for elastic contents] \label{elastic_th}
For caching elastic contents in a bipartite network in the above set up with $N \geq 2C$, the regret of any online caching policy $\pi$ is lower bounded as:
	\[R_T^\pi \geq d|\mathcal{J}|\sqrt{\frac{CT}{2\pi}} - \Theta(\frac{1}{\sqrt{T}}), ~~\forall T \geq 1.\]
\end{theorem}
\end{framed}
\paragraph{Proof outline:} In this proof, we construct a \emph{common} randomized file request sequence $\{\bm{X}_t\}_{t\geq 1}$, which is identical for each user. In other words, all users request the same random file at each slot. Thus, unlike most other applications of the probabilistic method, which usually proceeds with i.i.d. random variables, we consider a set of mutually \emph{dependent} file request sequence. The expected reward accrued by any caching policy is then obtained by using the statistical symmetry of the file requests and the linearity of expectation. Finally, the static optimal caching configuration is identified, and the reward accrued by the optimal stationary policy is lower bounded by appealing to Lemma \ref{max_load}. Combining the above two results yields the regret lower bound. Please refer to Section \ref{reg_bd_elastic} for the complete proof of Theorem \ref{elastic_th}. 

The following theorem gives regret lower bound for caching inelastic content in a bipartite network.

\begin{framed}
\begin{theorem}[Lower bound for inelastic contents] \label{inelastic_th}
For caching inelastic contents in a bipartite network in the above set up with $N \geq 2C|\mathcal{J}|$, the regret of any online caching policy $\pi$ is lower bounded as:
	\[R_T^\pi \geq d\sqrt{\frac{|\mathcal{J}|CT}{2\pi}} - \Theta(\frac{1}{\sqrt{T}}), ~~\forall T \geq 1. \]
\end{theorem}
\end{framed}
\paragraph{Proof outline:} The principal difficulty in extending the argument from the proof of Theorem \ref{elastic_th} to the inelastic case is the presence of non-linearity in the reward function \eqref{inelastic_content} in the form of $\min(\cdot, \cdot)$ operator. As a result, it becomes difficult to analyze the expected reward accrued by the optimal stationary caching configuration. To get around this obstacle, we lower bound the reward of the optimal caching configuration with the help of a carefully constructed sub-optimal caching configuration $\bm{Y}_\perp$. Interestingly, under the caching configuration $\bm{Y}_\perp$, the non-linearity of the reward function vanishes, which leads to tractable analysis. Similar to the proof of Theorem \ref{elastic_th}, this proof also uses an identical (\emph{i.e.,} dependent) file request sequence across all users. Please refer to Section \ref{reg_bd_inelastic} for the complete proof of Theorem \ref{inelastic_th}.  
\paragraph{Tightness:} Comparing the regret upper bounds in Theorems \ref{achievability} and \ref{elastic_uncoded} with the lower bounds in Theorems \ref{elastic_th} and \ref{inelastic_th}, we see that, for elastic contents, the OGA and \textsf{FTPL} network caching policies are regret-optimal up to a constant and poly-log factors respectively. For inelastic contents, the OGA policy is regret-optimal up to a factor of  $O(\sqrt{|\mathcal{J}|})$. 

\section{Experiments} \label{experiments}
\begin{figure*}
       \centering
%
\begin{subfigure}[b]{0.31\textwidth}
\begin{overpic}[width=\textwidth]{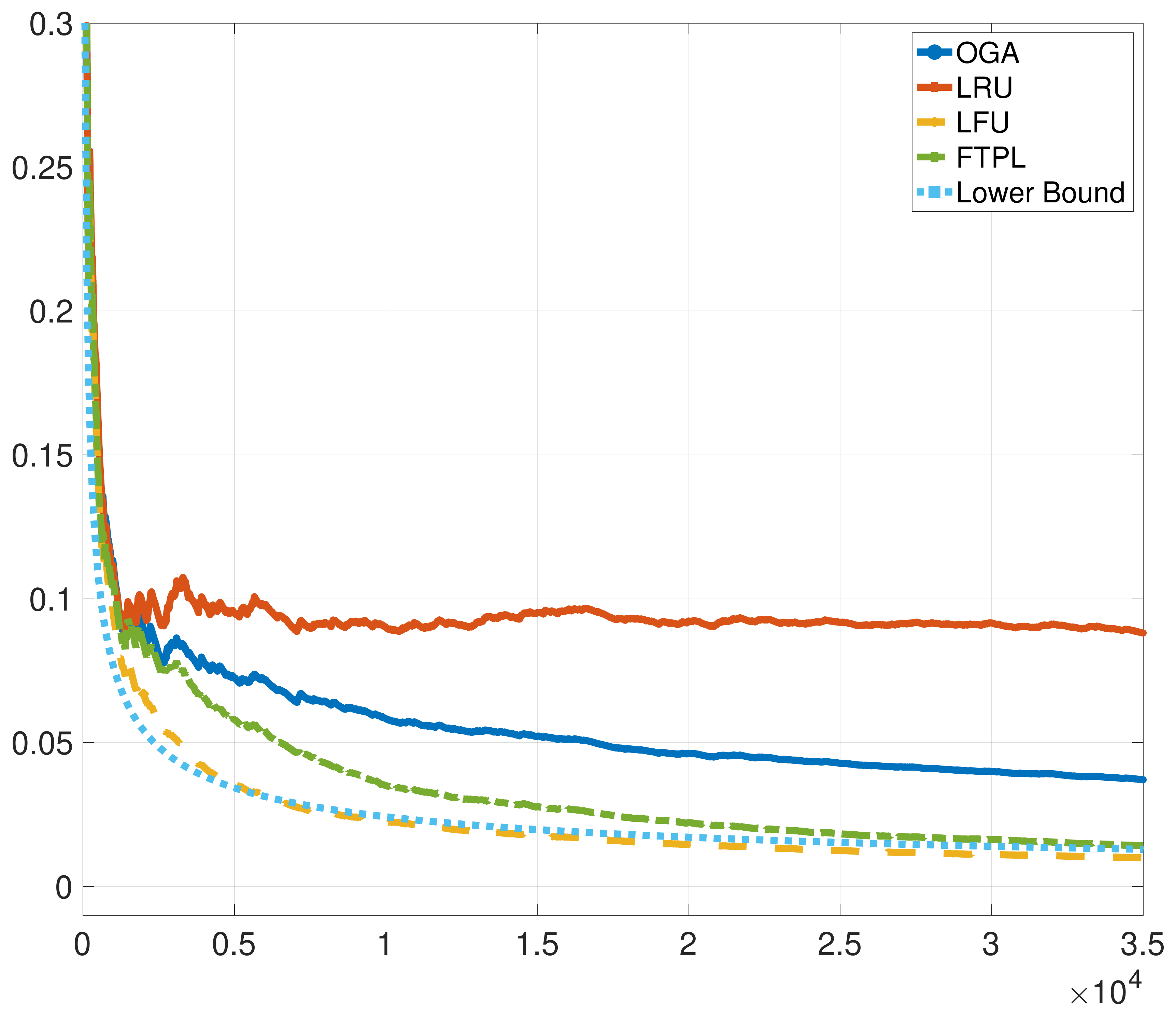}
\put(55,0){\footnotesize{$T$}}
\put(-3,40){\footnotesize{$\frac{R_T}{T}$}}
\end{overpic}

\end{subfigure}
\begin{subfigure}[b]{0.31\textwidth}
\begin{overpic}[width=\textwidth]{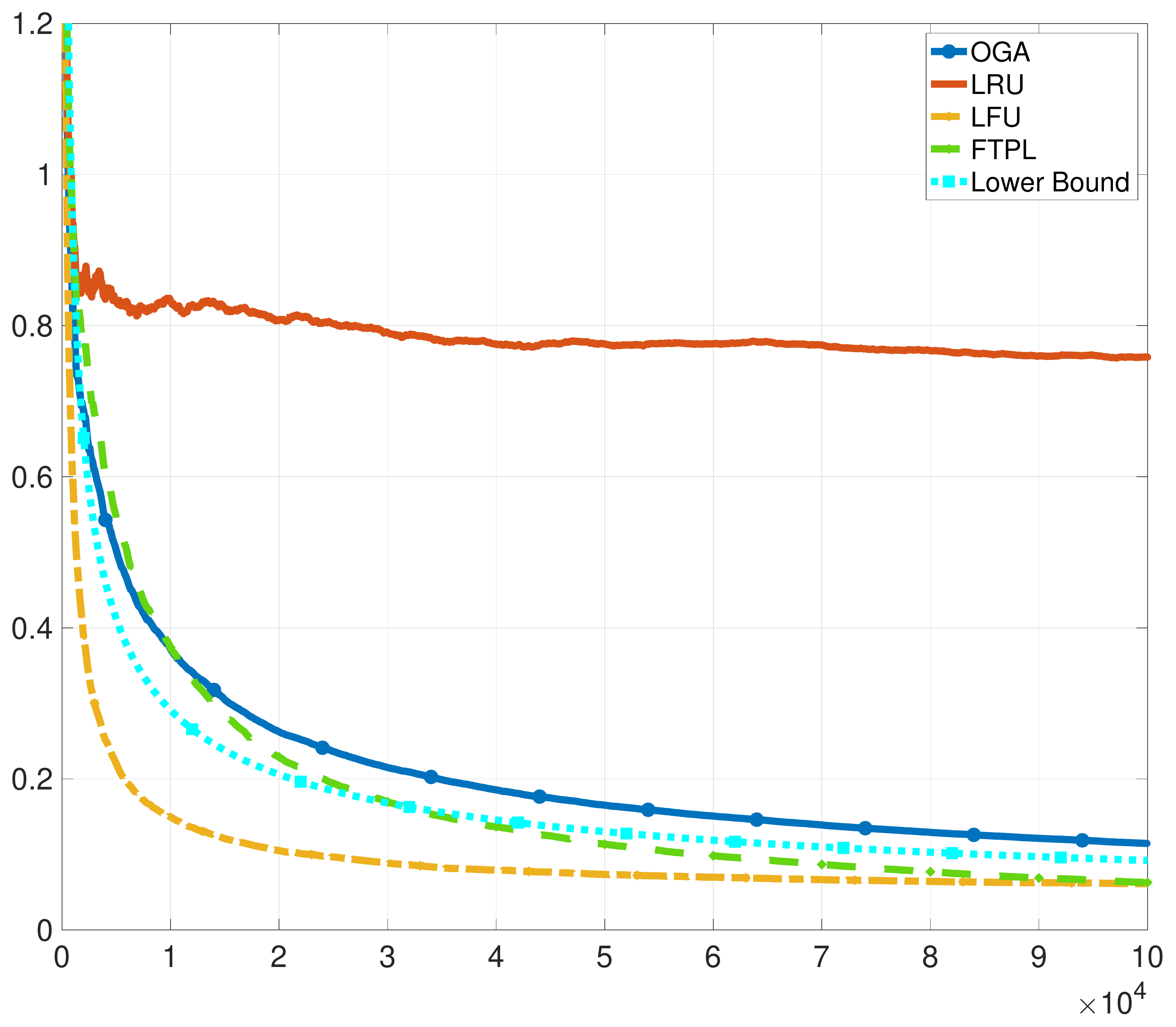}
\put(55,0){\footnotesize{$T$}}
\put(-2,53){\footnotesize{$\frac{R_T}{T}$}}
\end{overpic}
\end{subfigure}
%
\begin{subfigure}[b]{0.31\textwidth}
\begin{overpic}[width=\textwidth]{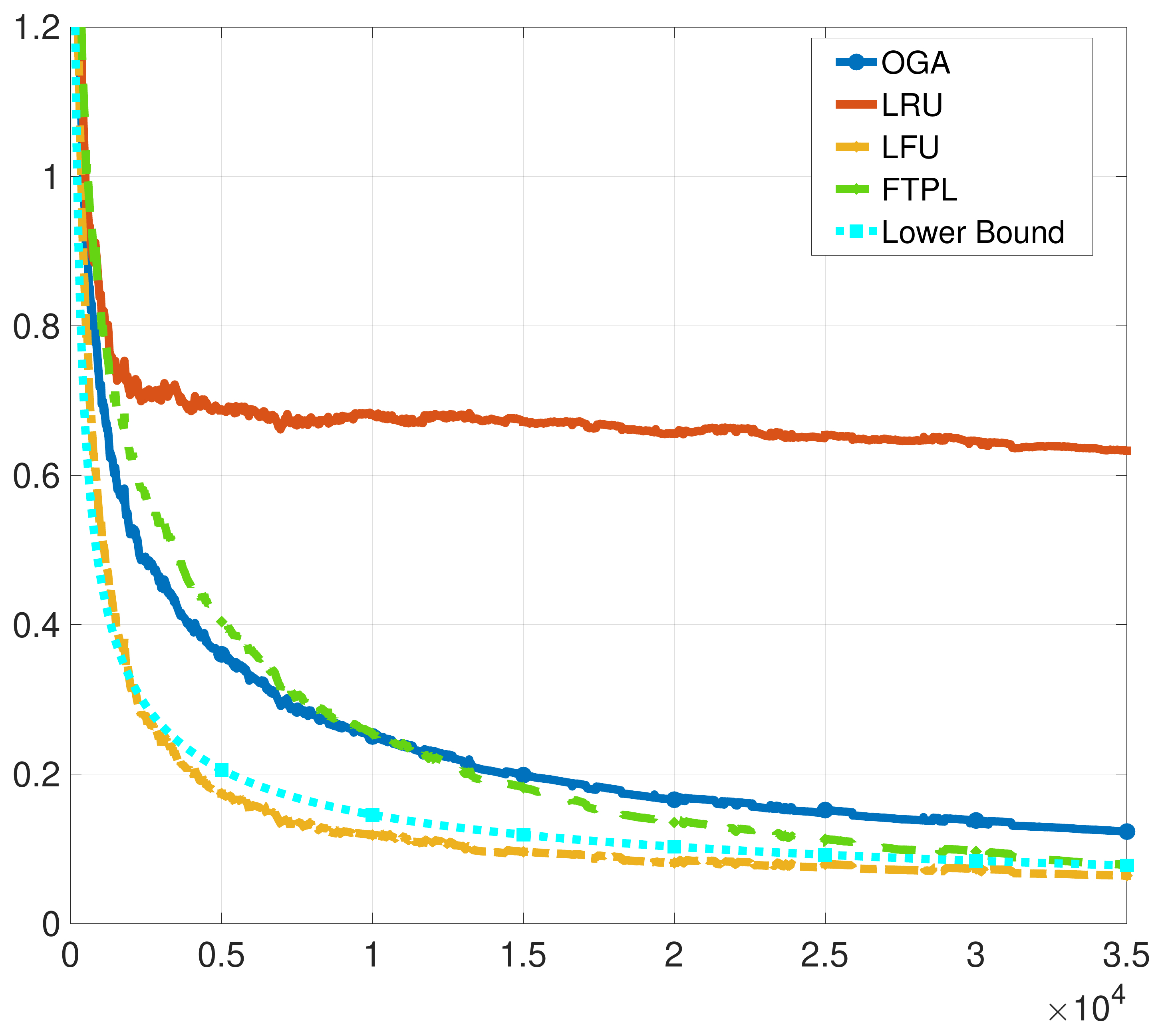}
\put(-2,53){\footnotesize{$\frac{R_T}{T}$}}
\put(55,0){\footnotesize{$T$}}
\end{overpic}
\end{subfigure}
\caption{\footnotesize{Comparison among the caching policies in terms of time-averaged regret $\frac{R_T}{T}$ for (a) single cache (b) bipartite caching network with elastic contents, and (c) bipartite caching network with inelastic contents. The bipartite network has $10$ users and $4$ caches each connected to $3$ users. The capacity of each cache is chosen to be $1$ per cent of the library size in all cases.}}
\label{regret_plot}
    \end{figure*}
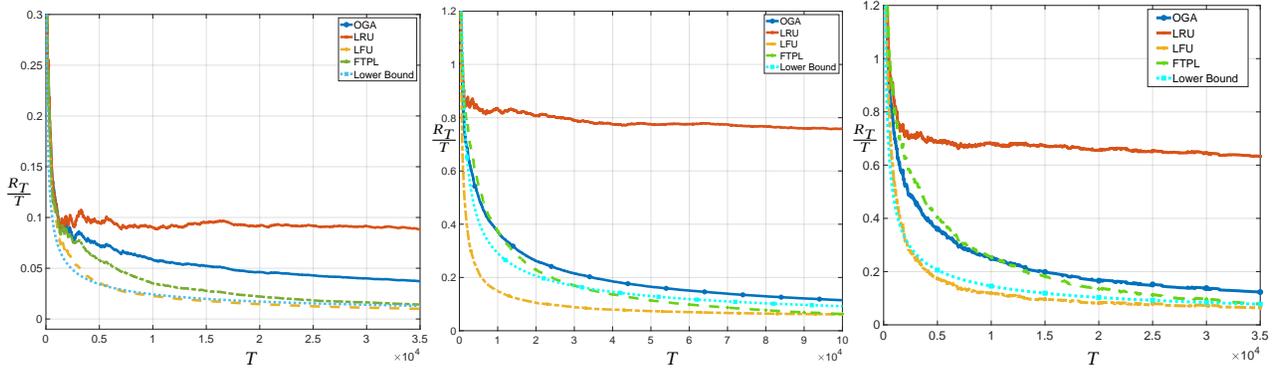

\emph{Dataset description:} In this section, we compare the performance of the existing caching policies with the proposed \textsf{FTPL} policy  using a popular and stable benchmark - \textsf{MovieLens $1$ M dataset} \cite{movieLens, harper2015movielens}. This dataset contains $\sim 1$ M  ratings for $N \sim 3700$ movies, along with the timestamps of the ratings. The ratings were given by $\sim 6000$ unique users.  For our experiments, we assume that the users request movies from a CDN (such as Netflix) in the same chronological order as the recorded timestamps of the ratings. A histogram of the movie request frequencies by all users together is shown in Fig. \ref{file_request_histogram} of the Appendix. 

\emph{Experimental Setup:} Following the standard industry practice, the cache capacity $C$ of each cache is set to be a fixed $\alpha$ fraction of the total library size $N$, where we take $\alpha=0.01$. For the bipartite caching scenario, we assume that a total of $|\mathcal{I}|=10$ users are connected to $|\mathcal{J}|=4$ caches. Each cache has in-degree $d=3$. The first cache is connected to the users $1,2$ and $3$, the second cache is connected to the users $4,5$ and $6$, the third cache is connected to the users $7,8$ and $9$, and the fourth cache is connected to the users $1,3$ and $10$. The entire dataset with $\sim 1M$ entries is uniformly divided into $10$ disjoint blocks. Each user is allocated one block of the dataset. It is assumed that each user makes requests serially from its allocated block in the  chronological order. 

\emph{Results:} The time-averaged regrets for different caching policies in the single cache setting and the bipartite network setting for both elastic and inelastic contents are plotted in Figure \ref{regret_plot} (see the following page). From the plots in Figures \ref{regret_plot} (a), 3 (b), and 3 (c), we conclude that the \textsf{FTPL} and \textsf{LFU}  policies have the best performance in terms of average regret uniformly in all cases, and there is hardly any noticeable difference between their performance for large enough $T$. These two policies also perform very close to the theoretical lower bounds (corresponding to the worst-case request sequence). On the other hand, we find that the LRU policy has the worst performance, followed by OGA, which performs only marginally better. It is quite surprising to find that the uncoded caching policy \textsf{FTPL} outperforms the coded caching policy \textsf{OGA} in all scenarios. Figure \ref{C_regret_plot} in the Appendix shows the variation of the average regret as the cache capacity is increased from $1$ percent to $10$ percent of the library size for a fixed time $T=5\times 10^4$. These plots confirm that the regret increases with the cache size. However, we find that the increase in the regret is smallest for the \textsf{LFU} and \textsf{FTPL} policies followed by the \textsf{OGA} and the \textsf{LRU} policy.

\section{Conclusion and future work} \label{conclusion}
In this paper, we obtain tight sub-linear regret lower bounds for the online caching problem for single caches and bipartite caching networks. In the process, we derive a key technical result on the \emph{balls-into-bins} problem and utilize the result in deriving all our lower bounds. We also propose a new randomized caching policy, called \textsf{FTPL}, which is shown to be both sound in theory and superior in practice. We envision the following future research directions stemming from this work:
 (1) As an immediate follow-up, it will be interesting to narrow down the $O(\sqrt{|\mathcal{J}|})$ gap between the lower and upper regret bounds for inelastic contents in a bipartite caching network. Moreover, obtaining a regret guarantee for the \textsf{FTPL} policy for bipartite networks with inelastic contents would be nice. 
 (2) We defined the reward functions primarily with the online performance of the caching policies (\emph{i.e.}, hit rates) in mind. In particular, our reward definitions do not take into account the system cost associated with cache replacements at every slot.  
Hence, it would be interesting to design an uncoded caching policy, which makes incremental changes to the caching configuration at every slot, yet matches the sublinear regret lower bounds. 
(3) A variation of the caching problem arises in the context of inventory management where, instead of digital files, physical commodities are stored in the caches (\emph{e.g.,} retail stores). Requests for the commodities arrive sequentially. The requested commodities, which are currently present in the cache, are immediately removed from the cache (\emph{e.g.,} sold). Hence, unlike in our setting, there is no scope of ``coding'', and it would make sense to cache multiple copies of the same commodity at the same slot, subject to the cache capacity constraints (c.f., Eqn.\ \eqref{admissible}) . The performance of \textsf{FTPL}-like randomized caching policies will be exciting to investigate in this setup. 
(4) Finally, it would also be interesting to go beyond the single-hop setting of bipartite caching networks and design a regret-optimal joint routing and caching policy for multi-hop CDNs \cite{liu2019joint}.

\section{Proof of the results}

\subsection{Proof of Lemma \ref{max_load}}
\label{max_load_proof}
\begin{figure}
\centering
\begin{overpic}[width=0.45\textwidth]{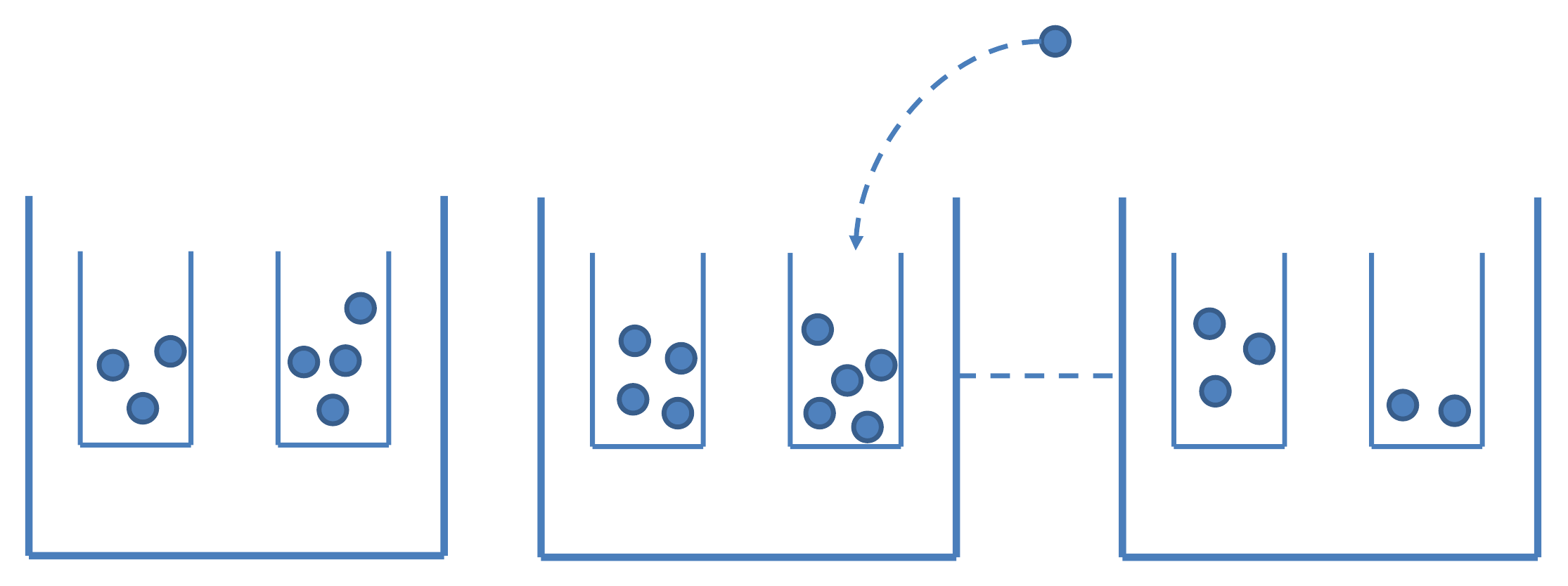}
\end{overpic}
\put(-213,12){\footnotesize{bin $1$}}
\put(-185,12){\footnotesize{bin $2$}}
\put(-140,12){\footnotesize{bin $3$}}
\put(-110,12){\footnotesize{bin $4$}}
\put(-53,14){\footnotesize{bin}}
\put(-58,6){\footnotesize{$2C-1$}}
\put(-28,12){\footnotesize{bin $2C$}}
\put(-210,-10){\footnotesize{Super bin $1$}}
\put(-135,-10){\footnotesize{Super bin $2$}}
\put(-47,-10){\footnotesize{Super bin $C$}}

\caption{\small {Illustrating the construction of Super bins}}
\label{superbins_fig}
\end{figure}
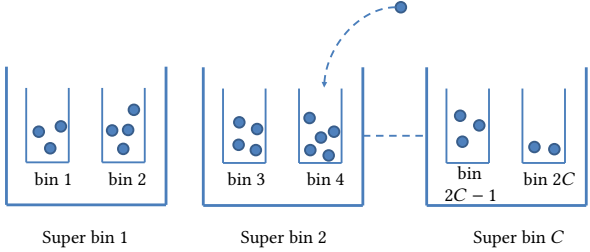

	We index the bins sequentially as $1,2,\ldots, 2C$. Next, we logically combine every two consecutive bins $\{(2i-1, 2i)\}, 1\leq i \leq C,$ to obtain $C$ \emph{Super bins} (See Figure  \ref{superbins_fig}). Let us denote the (random) number of balls in the $i$\textsuperscript{th} super bin by $X_i, j=1,2,\ldots, C.$ Conditioned on the r.v. $X_i$, the number of balls in the corresponding bins: $ 2i-1$ and $2i$ are jointly distributed as $(Z, X_i-Z)$, where $Z$ is a binomial random variable with parameter $(X_i, \frac{1}{2})$. Let $H_i$ denote the maximum number of balls between the corresponding bins $ 2i-1 $ and $2i$. Then, as shown in the proof Theorem \ref{reg_lb_spl}, when $X_i>0$:
	\begin{eqnarray} \label{lb2}
	\mathbb{E}(H_i|X_i) \geq  \frac{X_i}{2}+ \sqrt{\frac{X_i}{2\pi}}- \frac{1}{2\sqrt{2\pi X_i}},~ \forall 1\leq i\leq C.~
	\end{eqnarray}
Since $M_{C} \geq \sum_{i=1}^{C} H_i,$ 	we have
\begin{eqnarray}
&&\mathbb{E}(M_{C}) \nonumber \\
&\geq& \mathbb{E} \bigg(\sum_{i=1}^C H_i\bigg)
= \sum_{i=1}^C \mathbb{E}(H_i) 
\stackrel{(a)}{=} C \mathbb{E}(H_1)
\stackrel{(b)}{=} C \mathbb{E}\big(H_1 \mathds{1}(X_1>0)\big) \nonumber  \\
&\stackrel{(c)}=& C\mathbb{E}\mathbb{E}(H_1\mathds{1}(X_1>0)|X_1)\nonumber  \\
&\stackrel{(d)}{\geq}& C\mathbb{E}\bigg( \frac{X_1\mathds{1}(X_1>0)}{2}+ \sqrt{\frac{X_1}{2\pi}}\mathds{1}(X_1>0)-\frac{\mathds{1}(X_1>0)}{2\sqrt{2\pi X_1}}\bigg)\nonumber  \\
&\stackrel{(e)}{=}& \frac{C}{2}\mathbb{E}(X_1) + \frac{C}{\sqrt{2\pi}} \mathbb{E}(\sqrt{X_1})- \frac{C}{2\sqrt{2\pi }} \mathbb{E}\bigg(\frac{\mathds{1}(X_1>0)}{\sqrt{X_1}}\bigg), \label{M_bd}
\end{eqnarray}
where the equality (a) follows from the fact that the random variables $\{H_i\}_{i=1}^C$ have identical distribution. For equation (b), we write 
\[ H_1 = H_1 \mathds{1}(X_1=0) + H_1 \mathds{1}(X_1 >0). \]
Now, observe that if $X_1=0$ then $H_1=0$ a.s. Hence, almost surely, we have $H_1=H_1\mathds{1}(X_1 >0)$.
  The equation (c) follows from the tower property of conditional expectation, the inequality (d) follows from the bound \eqref{lb2}, and the equality (e) follows from the facts that $X_1 = X_1 \mathds{1}(X_1>0), \sqrt{X_1} = \sqrt{X_1}\mathds{1}(X_1>0).$ The Lemma now follows by using the bounds on moments of the binomial distribution as computed next.

 \paragraph{Bounding the Expectations in Eqn.\ \eqref{M_bd}:}
 
For bounding the middle term in Eqn.\ \eqref{M_bd}, consider the factorization \cite{stack_of}:
\begin{eqnarray*} 
\sqrt{x} - \bigg( 1 + \frac{x-1}{2} - \frac{(x-1)^2}{2}\bigg) = \frac{\sqrt{x}}{2}(\sqrt{x}-1)^2(\sqrt{x}+2). 	
\end{eqnarray*}
 The RHS is non-negative for any $x\geq 0$. Thus, we have the following algebraic inequality:
 \begin{eqnarray*}
 	\sqrt{x} \geq 1 + \frac{x-1}{2} - \frac{(x-1)^2}{2}, ~~ \forall ~ x\geq 0.  
 \end{eqnarray*}

 Replacing the variable $x$ with the random variable $\frac{X_1}{\mathbb{E}(X_1)}$ point wise, we have almost surely,
 \begin{eqnarray*}
 	\sqrt{\frac{X_1}{\mathbb{E}(X_1)}} \geq 1 + \frac{\frac{X_1}{\mathbb{E}(X_1)}-1}{2} - \frac{(\frac{X_1}{\mathbb{E}(X_1)}-1)^2}{2}.
 \end{eqnarray*}
Taking expectation of both sides, the above yields 
 \begin{eqnarray}\label{final_ineq}
 \mathbb{E}(\sqrt{X_1}) \geq \sqrt{\mathbb{E}(X_1)}\bigg( 1 - \frac{\textsf{Var}(X_1)}{2(\mathbb{E}(X_1))^2}\bigg).	
 \end{eqnarray}
Finally, recall that $X_1 \sim \textsf{Binom}(T,\frac{1}{C})$. Hence, $\mathbb{E}(X_1)= \frac{T}{C}$ and $\textsf{Var}(X_1) = T \frac{1}{C}(1-\frac{1}{C}) \leq \frac{T}{C}$.
Using this, Eqn.\ \eqref{final_ineq} yields the following lower bound 
\begin{eqnarray} \label{lb_mid_term}
	\mathbb{E}(\sqrt{X_1}) \geq \sqrt{\frac{T}{C}} - \frac{1}{2} \sqrt{\frac{C}{T}}. 
\end{eqnarray}
 For bounding the last term in Eqn.\ \eqref{M_bd}, we write
\begin{eqnarray*}
 \frac{\mathds{1}(X_1>0)}{\sqrt{X_1}} \leq 1 \mathds{1}\big(X_1 \leq \frac{T}{2C}\big) + \sqrt{\frac{2C}{T}}.
 \end{eqnarray*}
 Recall that $X_1 \sim \textsf{Binom}(T, \frac{1}{C}).$ Taking expectation of both sides of the above inequality, we have
\[\mathbb{E}\bigg(\frac{\mathds{1}(X_1>0)}{\sqrt{X_1}}\bigg) \leq \mathbb{P}(X_1 \leq \frac{1}{2}\mathbb{E}(X_1)) + \sqrt{\frac{2C}{T}}. \]
The probability term in the above expression may be bounded using Chebyshev's inequality as follows:
\begin{eqnarray*}
\mathbb{P}\big(X_1 \leq \frac{1}{2}\mathbb{E}(X_1)\big) &=& \mathbb{P}\big(X_1 - \mathbb{E}(X_1)\leq -\frac{1}{2}\mathbb{E}(X_1)\big)	\\
&\leq & \mathbb{P}\big(|X_1 - \mathbb{E}(X_1)|\geq \frac{1}{2}\mathbb{E}(X_1)\big)\\
&\leq & \frac{4\textsf{Var}(X_1)}{(\mathbb{E}(X_1))^2}= 4\frac{C-1}{T}
\end{eqnarray*}
Taking the above bounds together, we obtain 
\begin{eqnarray} \label{bd_last_term}
	\mathbb{E}\bigg(\frac{\mathds{1}(X_1>0)}{\sqrt{X_1}}\bigg) \leq \sqrt{\frac{2C}{T}} + \frac{4C}{T}.
\end{eqnarray}

Finally, combining Eqns.\ \eqref{M_bd}, \eqref{lb_mid_term}, and \eqref{bd_last_term} together, we obtain 
\begin{eqnarray*}
\mathbb{E}(M_C) \geq \frac{T}{2} + \sqrt{\frac{CT}{2\pi}} - \frac{(\sqrt{2}+1)C^{3/2}}{2\sqrt{2 \pi T}}- \sqrt{\frac{2}{\pi}}\frac{C^2}{T}. ~~~\blacksquare	
\end{eqnarray*}


\subsection{Proof of Theorem \ref{reg_lb_gnl}} \label{reg_lb_gnl_proof}
To lower bound the regret $R_T^\pi$ in equation \eqref{regret_def}, as in Section \ref{basic}, we consider a random file request sequence $\{\bm{X}_t\}_{t\geq 1}$, each sampled independently and uniformly at random from the set of first $2C$ unit vectors $\{\bm{e}_i, 1\leq i \leq 2C\}$ of dimension $N$ \footnote{Recall that, the unit vector $\bm{e}_i$ has one at its $i$\textsuperscript{th} coordinate and zeros everywhere else. }. In other words, at every time slot, the user independently requests a random file from the set of first $2C$ files uniformly at random.

The expected reward obtained by any caching policy, given by the second term in Eqn.\ \eqref{regret_seq}, is now easy to evaluate: 
\begin{eqnarray} \label{ex2}
\mathbb{E}\bigg(\sum_{t=1}^{T} \bm{Y}_t \cdot \bm{X}_t\bigg) \stackrel{(a)}{=} \sum_{t=1}^{T}\bm{Y}_t \cdot \big(\mathbb{E}\bm{X}_t\big)
= \frac{1}{2C}\sum_{t=1}^{T} \sum_{k=1}^{2C}\bm{Y}_{tk}  
 \stackrel{(b)}{\leq} \frac{T}{2}, 
\end{eqnarray}
where, in (a), we have used the fact that the caching decision $\bm{Y}_t$ made at time $t$ is independent of the incoming file request $\bm{X}_t$, and in (b), we have made use of the cache-capacity constraint \eqref{admissible}:
\[\sum_{k=1}^{2C}\bm{Y}_{tk} \leq \sum_{k=1}^{N} \bm{Y}_{tk} \leq C,\]
in addition to the fact that
\begin{eqnarray*} 
\mathbb{E}(X_{tk})= \begin{cases} \frac{1}{2C}, ~~ \forall 1 \leq k \leq  2C \\
0 ~~ \textrm{o.w.}
 \end{cases}
\end{eqnarray*}
Note that, for any given file request sequence $\{\bm{x}_t\}_{t=1}^{T}$, the optimal \emph{offline} stationary cache configuration vector is obtained by caching the most popular $C$ files. Hence, the optimal vector $\bm{y}^* \in [0,1]^N$, corresponding to the first term in Eqn.\ \eqref{regret_seq} is obtained by simply setting the coordinates corresponding to the \emph{maximum $C$ coordinates} of the $N$-dimensional vector $\sum_{t=1}^{T} \bm{x}_t$ to unity. 
Given the distribution of the file request sequence, it immediately follows that the reward accrued by the optimal stationary policy ${\bm{Y^*}}\cdot \sum_{t=1}^{T} \bm{X}_t$ is identically distributed to the total number of balls in the most heavily loaded $C$ bins when a total of $T$ balls are randomly thrown into $2C$ bins. Finally, invoking Lemma \ref{max_load}, we have 
\begin{eqnarray} \label{ex1}
\mathbb{E}\bigg({\bm{Y^*}} \cdot \sum_{t=1}^{T} \bm{X}_t\bigg) \geq \frac{T}{2}+ \sqrt{\frac{CT}{2\pi}} - \Theta(\frac{1}{\sqrt{T}}).
\end{eqnarray}
Combining equations \eqref{ex2} and \eqref{ex1}, we obtain the following lower bound for regret in the single cache setting: 
\begin{eqnarray} \label{reg_bd}
R_T&\geq & \mathbb{E}_{\{\bm{X}_t\}_{t=1}^{T}}\bigg({\bm{Y^*}}\cdot \sum_{t=1}^{T} \bm{X}_t - \sum_{t=1}^{T}\bm{Y}_t \cdot \bm{X}_t \bigg) \nonumber \\
&\geq&  \sqrt{\frac{CT}{2\pi}}- \Theta(\frac{1}{\sqrt{T}}).~~~\blacksquare \nonumber
\end{eqnarray}


\subsection{Proof of Theorem \ref{uc_achievability}} \label{uc_achievability_proof}
Our proof follows a similar line of arguments as the proof of Theorem 1 of \cite{cohen2015following}. However, we improve the regret upper bound by a factor of $O(\sqrt{C})$. This additional improvement results from making use of the constraint that only one file is requested at every slot. The notations of \cite{cohen2015following} are slightly altered in order to remain consistent throughout the paper. 

Let the set $\mathcal{Y}$ denote the set of all possible uncoded caching configuration in the single cache setting. Clearly, $|\mathcal{Y}| = \binom{N}{C}.$ Define the potential function
\[ \Phi_\eta(\bm{x})= \mathbb{E}_{\bm{\gamma}\sim \mathcal{N}(0,I)}\bigg[\max_{\bm{y}\in \mathcal{Y}} \langle \bm{y}, \bm{x}+ \eta \bm{\gamma} \rangle \bigg].\]
Also, denote the cumulative file request arrivals to the cache up to time $t-1$ by $\bm{X}_t = \sum_{\tau=1}^{t-1}\bm{x}_\tau$ with $\bm{X}_1=\bm{0}.$ Then, as shown in Eqn.\ (3) of \cite{cohen2015following}, the expected regret of the \textsf{FTPL} policy in Algorithm \ref{uncoded} with noise variance $\eta^2$ is upper bounded as \footnote{The signs are flipped as we are in the rewards maximization setting, as opposed to the loss minimization setting of \cite{cohen2015following}.}
\begin{eqnarray} \label{main_bd}
\mathbb{E}(R_T) \leq \Phi_\eta(\bm{X}_1)+ \frac{1}{2}\sum_{t=1}^T\langle \bm{x}_t, \nabla^2\Phi_\eta(\tilde{\bm{x}}_t)\bm{x}_t)\rangle,
\end{eqnarray}
for some $\tilde{\bm{x}}_t$ connecting the line segment $\bm{X}_t$ and $\bm{X}_{t+1}$. \\
Next, we bound each of the above two terms separately. The first term may be bounded in the same way as in \cite{cohen2015following}:
\begin{eqnarray*}
	\Phi_\eta(\bm{X}_1) \leq \eta \sqrt{2C\log \binom{N}{C}} \leq C\eta\sqrt{2\log N}, 
\end{eqnarray*}
where the last inequality follows from the fact that $\binom{N}{C} \leq N^C$. Since only one file is requested at every slot, the quadratic form above may be upper bounded as 
\begin{eqnarray} \label{quad_form1}
	\langle \bm{x}_t, \nabla^2\Phi_\eta(\tilde{\bm{x}_t})\bm{x}_t)\rangle \leq \max_{i, \bm{x}} \big(|\nabla^2\Phi_\eta(\bm{x})\big|)_{ii}.
\end{eqnarray}
Moreover, following Lemma 7 of \cite{abernethy2014online}, we have that
\begin{eqnarray*}
	(\nabla^2 \Phi_\eta(\bm{x})\big)_{ij} = \frac{1}{\eta} \mathbb{E}\bigg[\hat{y}(\tilde{\bm{x}}_t+\eta \bm{\gamma})_i\gamma_j\bigg],
\end{eqnarray*}
where $\hat{y}(z)\in \arg\max_{\bm{y} \in \mathcal{Y}}\langle \bm{y}, \bm{z} \rangle. $ Hence, using Jensen's inequality we have that
\begin{eqnarray}\label{hess_bd}
	(\big|\nabla^2\Phi_\eta(\bm{x})\big|)_{ii} \leq \frac{1}{\eta} \mathbb{E}\bigg[|\hat{y}(\tilde{\bm{x}}_t+\eta \bm{\gamma})_i||\gamma_i|\bigg] \stackrel{(a)}{\leq} \frac{1}{\eta}\mathbb{E}\big[|\gamma_i|\big] \stackrel{(b)}{=} \frac{1}{\eta}\sqrt{\frac{2}{\pi}},
\end{eqnarray}
where the inequality (a) follows from the fact that for all $\bm{y} \in \mathcal{Y}$, we have $y_i \in \{0,1\},$ and the equality (b) follows from the fact that $\gamma_i \sim \mathcal{N}(0,1).$ Hence, substituting the above bounds in Eqn.\ \eqref{main_bd}, we have the following upper bound on expected regret under the \textsf{FTPL} caching policy: 
\begin{eqnarray*}
	\mathbb{E}(R_T) \leq C\eta \sqrt{2 \log N}+ \frac{T}{\eta\sqrt{2\pi}}.
\end{eqnarray*}
Finally, choosing $\eta = \frac{1}{(4 \pi \log N)^{1/4}}\sqrt{\frac{T}{C}},$ yields the following regret upper bound for the \textsf{FTPL} policy:
\begin{eqnarray*}
	\mathbb{E}_{\{\bm{\gamma}_t\}_t}(R_T^{\textsf{FTPL}}) \leq  1.51(\log N)^{1/4}\sqrt{CT}. ~~~\blacksquare
\end{eqnarray*}

\subsection{Proof of Theorem \ref{achievability}} \label{achievability_proof}
From Eqn.\ \eqref{elastic_content}, we know that $q_{\textsf{elastic}}(\bm{x}, \bm{y})$ is linear (and hence, concave) in the cache-configuration vector $\bm{y}$. Moreover, since the pointwise minimum of linear functions is concave \cite{bertsekas}, it follows from Eqn.\ \eqref{inelastic_content} that the reward function $q_{\textsf{inelastic}}(\bm{x}, \bm{y})$ is also concave in the cache-configuration vector $\bm{y}$. To obtain a regret upper bound for the the OGA algorithm \eqref{bsa}, we appeal to Theorem 2 of \cite{paschos2019learning}, which states that with an appropriate choice of the step-size parameter $\eta$, 
\begin{eqnarray} \label{reg_ub}
	R_T^{\textsf{OGA}} \leq \textrm{diam}(Y_{\mathcal{J}})L \sqrt{T}, 	
\end{eqnarray}
where $\textrm{diam}(Y_{\mathcal{J}})$ denotes the Euclidean diameter \cite{rudin1964principles} of the set $Y_\mathcal{J}$ defined in \eqref{feasibility}, and $L$ is an upper-bound for the $2$-norm of the (super) gradient of the reward function. \\
For bounding the diameter, consider any two vectors $\bm{y}$ and $\bm{z}$ from the set $Y_{\mathcal J}$. We have 
\begin{eqnarray*}
|| \bm{y} - \bm{z}||_2^2 &=& \sum_{j \in \mathcal J} \sum_{f=1}^{N} (y^j_f - z^j_f)^2 \\
& \stackrel{(a)}{\leq} & \sum_{j \in \mathcal J} \sum_{f=1}^{N} |y^j_f - z^j_f|  \\
& \stackrel{(b)}{\leq}& \sum_{j \in \mathcal J} \sum_{f=1}^{N} y^j_f + \sum_{j \in \mathcal J} \sum_{f=1}^{N} z^j_f 
 \stackrel{(c)}{\leq} 2CJ, 	
\end{eqnarray*}
where the inequality (a) follows from the fact that $|y^j_f - z^j_f| \leq 1$, the inequality (b) follows from triangle inequality, and finally, inequality (c) follows from the cache-capacity constraints. Since, the above bound is valid for any two vectors in the set $Y_{\mathcal{J}}$, it follows that  
\begin{eqnarray} \label{diam_bd}
\textrm{diam}(Y_{\mathcal{J}}) \equiv \sup_{\bm{y}, \bm{z} \in Y_{\mathcal J}}|| \bm{y}- \bm{z}|| \leq \sqrt{2CJ}.
\end{eqnarray}
Next, we bound the norm of the supergradients of the reward functions. Recall,
\[q_{\textsf{elastic}}(\bm x, \bm y) = \sum_{i \in \mathcal{I}} \bm{x}^i \cdot \big(\sum_{j \in \partial^+(i)} \bm{y}^j \big).\] 
Hence, we have the following expression for the supergradient of the objective function:
\begin{eqnarray} \label{sup_grad_expr}
 \bigg(\nabla_{\bm{y}} q_{\textsf{elastic}} (\bm{x}, \bm{y})\bigg)^j_f = \sum_{i \in \partial^-(j)} x^i_f. 
 \end{eqnarray}
Thus, 
\begin{eqnarray} \label{norm_bound}
 ||\nabla_{\bm{y}} q_{\textsf{elastic}} (\bm{x}, \bm{y})||_2^2= \sum_{j \in \mathcal{J}} \sum_f  \big(\sum_{i \in \partial^-(j)} x^i_f\big)^2. 
 \end{eqnarray}
On the other hand, since $\bm{x} \geq \bm{0}$, it follows that the vector $\nabla_{\bm{y}} q_{\textsf{elastic}} (\bm{x}, \bm{y})$ can be taken to be a supergradient of the concave function $q_{\textsf{inelastic}} (\bm{x}, \bm{y})$ w.r.t. the argument $\bm{y}.$ Thus, to obtain the regret upper bound, it only remains to upper bound the RHS of Eqn.\ \eqref{norm_bound}. We have:
\begin{eqnarray*}
	\big(\sum_{i \in \partial^-(j)} x^i_f\big)^2 \stackrel{(a)}{\leq} d \sum_{i \in \partial^-(j)} (x^i_f)^2 \stackrel{(b)}{=} d\sum_{i \in \partial^-(j)} x^i_f,
\end{eqnarray*}
where the inequality (a) follows from Cauchy-Schwartz inequality and (b) follows from the fact that each $x_f^i$'s are either zero or one. Hence, the RHS of Eqn.\ \eqref{norm_bound} is upper bounded as follows:
\begin{eqnarray*}
	\sum_{j \in \mathcal{J}} \sum_f  \big(\sum_{i \in \partial^-(j)} x^i_f\big)^2 &\leq& d \sum_{j \in \mathcal{J}} \sum_f \sum_{i \in \partial^-(j)} x^i_f \\
	& = & d \sum_{j \in \mathcal{J}}\sum_{i \in \partial^-(j)}\sum_f x^i_f \\
	&\stackrel{(a)}{\leq}& d \sum_{j \in \mathcal{J}}\sum_{i \in \partial^-(j)} 	1 = d^2|\mathcal{J}|,
\end{eqnarray*}
where (a) follows from the facts that each user can request at most one file per slot, and the network is right $d$-regular. Hence, for both elastic and inelastic reward functions, the 2-norm of the supergradients is upper bounded as 
\begin{eqnarray}\label{norm_bd_final}
  L \leq d\sqrt{|\mathcal{J}|}. 
\end{eqnarray} 
Hence, combining Eqns. \eqref{reg_ub}, \eqref{diam_bd}, and \eqref{norm_bd_final}, we have that 
\[ R_T^{\textsf{OGA}} \leq d|\mathcal{J}|\sqrt{2CT}. ~~~\blacksquare\]

\subsection{Proof of Theorem \ref{elastic_th}} \label{reg_bd_elastic}
We prove a slightly general result without assuming the bipartite caching network to be right $d$-regular. Let $\bm{x}_t^{i}$ be the file request vector from the $i$\textsuperscript{th} user, and $\bm{y}_t^{j}$ be the cache configuration vector of the $j$\textsuperscript{th} cache selected by the policy $\pi$ at time $t$, where $i \in \mathcal{I}, j \in \mathcal{J}$. Hence, due to the elastic nature of the content, the total reward $G^\pi_T$ accrued by the caching policy $\pi$ is given by Eqn.\ \eqref{elastic_content}: 
\[ G_T^\pi = \sum_{t=1}^{T} \sum_{i \in \mathcal{I}} \bm{x}^i_t \cdot \sum_{j\in \partial^+(i)}\bm{y}_t^j.\]
Recall that, while computing the regret of a policy $\pi$ via Eqn.\ \eqref{regret_seq}, we compare the reward accrued by the policy $\pi$ to that of a stationary caching policy $\pi^*$ equipped with the hindsight knowledge (\emph{i.e.,} $\pi^*$ knows the entire file request sequence $\{\bm{x}_t\}_{1}^{T}$ in advance, \emph{viz.,} Eqns.\ \eqref{regret_seq}- \eqref{regret_def1}). Let $\bm{y}^j_*$ be the optimal stationary cache configuration vector at the server $j \in \mathcal{J}$ set by the policy $\pi^*$. Then the reward accrued by the stationary caching policy $\pi^*$ is given by: 
 \[ G_T^{\pi^{*}} = \sum_{t=1}^{T} \sum_{i \in \mathcal{I}} \bm{x}^i_t \cdot \sum_{j\in \partial^+(i)}\bm{y}_*^j.\]
Rearranging the terms, we have
\begin{eqnarray} \label{elastic_lb0}
 G_T^{\pi^*} = \sum_{j \in \mathcal{J}} \bm{y}_*^j\cdot \big( \sum_{i \in \partial^-(j)}\sum_{t=1}^{T} \bm{x}_t^i\big).
 \end{eqnarray}
Since the $j$\textsuperscript{th} cache has a total capacity of $C$, from the above equation, it is clear that the optimal stationary configuration $\bm{y}^j_*$ corresponds to caching the most popular $C$ files requested by all of $j$'s in-neighbours taken together. 
Following the probabilistic method described in Section \ref{basic}, we now construct a randomized file request sequence which are \emph{identical} for each user, \emph{i.e.}, all users request the \emph{same} random file at every time slot. At time slot $t$, the file request from the $i$\textsuperscript{th} user is given by the vector $\bm{X}_t^{i}=\bm{X}_t$, where the vector $\bm{X}_t$
 is sampled independently and uniformly at random from the set of first $2C$ unit vectors $\{\bm{e}_i \in \mathbb{R}^N, 1\leq i \leq 2C\}$. 
With this set up, the expected reward accrued by the policy $\pi$ (\emph{i.e.,} the second term in Eqn.\ \eqref{regret_seq}) may be evaluated as: 
\begin{eqnarray*} \label{elastic_lb1}
\mathbb{E}\bigg( \sum_{t=1}^{T} \sum_{i \in \mathcal{I}} \bm{X}_t^{i} \cdot \sum_{j\in \partial^+(i)}\bm{Y}_t^j\bigg)
&=& \sum_{t=1}^{T} \sum_{i \in \mathcal{I}} \sum_{j \in \partial^+(i)} \sum_{f=1}^{N} \mathbb{E}(X_{tf})Y^j_{tf} \nonumber\\
&\stackrel{(a)}{=}& \frac{1}{2C} 	\sum_{t=1}^{T} \sum_{i \in \mathcal{I}} \sum_{j \in \partial^+(i)} \sum_{f=1}^{2C} Y^j_{tf} \nonumber \\
&\stackrel{(b)}{\leq} & \frac{1}{2C} \sum_{t=1}^{T} \sum_{i \in \mathcal{I}} \sum_{j \in \partial^+(i)} C 
\stackrel{(c)}{=}\frac{T}{2}\sum_j d_j ,
\end{eqnarray*}
where the equation (a) uses the fact that the random variables $\bm{X}_t$ and $\bm{Y}_t$ are independent and $\mathbb{E}(X_{tf})=\frac{1}{2C}, \forall t,f$. The inequality (b) follows from the fact that each cache has capacity $C$, and the equality (c) follows from interchanging the order of the summations. \\
From Eqn.\ \eqref{elastic_lb0}, we can express the expected total reward $G^{\pi^*}_T$ accrued by the stationary policy $\pi^*$ as: 
\begin{eqnarray*}
\mathbb{E}(G^{\pi^*}_T) = \sum_{j \in \mathcal{J}}  \mathbb{E}(M^j_C),
\end{eqnarray*}
where $M^j_C$ denotes the sum of the largest $C$ coordinates of the following $N$-dimensional random vector:
\[\sum_{t=1}^{T} \sum_{i \in \delta^-(j)} \bm{X}_t^{i}= d_j \sum_{t=1}^{T} \bm{X}_t.\] 
Thus, it follows that the random variable $M^j_C$ is statistically identical to $d_j$ times the total number of balls in the most occupied $C$ bins when $T$ number of balls are thrown independently and uniformly at random into $2C$ bins. Finally, appealing to Lemma \ref{max_load}, we have 
\begin{eqnarray} \label{elastic_lb2}
\mathbb{E}(M^j_C) &\geq& d_j\bigg(\frac{T}{2} + \sqrt{\frac{CT}{2\pi}} - \Theta(\frac{1}{\sqrt{T}})\bigg).
\end{eqnarray}
Finally, combining the above, we have the following regret lower bound for caching elastic contents in a bipartite caching network:
\begin{eqnarray*}
R^\pi_T \geq \sqrt{\frac{CT}{2\pi}}\bigg(\sum_{j \in \mathcal{J}} d_j\bigg) - \Theta(\frac{1}{\sqrt{T}})= d|\mathcal{J}|\sqrt{\frac{CT}{2\pi}} - \Theta(\frac{1}{\sqrt{T}}). ~~\blacksquare
\end{eqnarray*}
\subsection{Proof of Theorem \ref{inelastic_th}} \label{reg_bd_inelastic}
We begin our analysis with the following two observations:\\
\begin{enumerate}
\item  From the definition of the reward functions \eqref{elastic_content} and \eqref{inelastic_content}, in general, we have, 
\begin{eqnarray} \label{obs_1}
  q_{\textsf{inelastic}}(\bm{x}, \bm{y}) \leq q_{\textsf{elastic}}(\bm{x}, \bm{y}), ~~\forall \bm{x}, \bm{y}. 
 \end{eqnarray}
\item In the special case, when different servers $j \in \mathcal{J}$ cache \emph{different} items, \emph{i.e.,} $\bm{y}^i \cdot \bm{y}^j=0, \forall i\neq j$, we have 
\begin{eqnarray} \label{obs_2}
 q_{\textsf{inelastic}}(\bm{x}, \bm{y}) = q_{\textsf{elastic}}(\bm{x}, \bm{y}), ~~\forall \bm{x}, \bm{y}. 
 \end{eqnarray}
 \end{enumerate}
Eqn.\ \eqref{obs_2} follows from the fact that in the product $ \bm{x}_t^i \cdot \big(\sum_{j \in \partial^+(i)} \bm{y}_t^j\big)= \sum_{j \in \partial^+(i)}\bm{x}_t^i \cdot \bm{y}_t^j$, only one of the dot-product terms could be strictly positive, as different servers cache different items and the vector $\bm{x}_t^{i}$ has only one positive component. Moreover, the value of each dot product is at most one, as each server stores at most one copy of each file.  \\
With these two observations at hand, we now relate our argument below to the proof of Theorem \ref{elastic_th} to obtain a regret lower bound for inelastic contents. Similar to the proof of Theorem \ref{elastic_th}, we construct a randomized and \emph{identical} file request sequence $\{\bm{X}_t^{i}\equiv \bm{X}_t\}_{t\geq 1}$ for each user, where the random variable $\bm{X}_t$ is sampled independently and uniformly at random from the set of \emph{first} $2C|\mathcal{J}|$ unit vectors $\{\bm{e}_i \in \mathbb{R}^N, 1 \leq i \leq 2C|\mathcal{J}|\}$.  Hence, using observation (1) and working similarly as in Eqn.\ \eqref{elastic_lb1}, the expected total reward $G_T^\pi$ accrued by any online caching policy may be upper bounded as 
\begin{eqnarray}\label{inelastic_lb1}
	G_T^\pi \leq \mathbb{E}\bigg( \sum_{t=1}^{T} \sum_{i \in \mathcal{I}} \bm{X}_t^{i} \cdot \sum_{j\in \partial^+(i)}\bm{Y}_t^j\bigg)
	\leq  \frac{T}{2|\mathcal{J}|}\sum_{j \in \mathcal{J}} d_j 	= \frac{Td}{2}.
\end{eqnarray}
Note that, unlike the elastic setting, due to the presence of the $\min$ operator in the reward function \eqref{inelastic_content}, obtaining an optimal cache configuration vector $\bm{Y}^*$ is non-trivial in the inelastic setting. However, since we only require a lower bound to the total expected reward accrued by the optimal static policy $\pi^*$, any suitably constructed sub-optimal caching configuration will serve the purpose, provided we can evaluate its reward. 
Towards this end, in the following, we construct an $N|\mathcal{J}|$ dimensional stationary cache-configuration vector $\bm{Y}_\perp$ (obtained by stacking the $N$-dimensional cache-configuration vectors $\bm{Y}_\perp^j$ together $\forall j \in \mathcal{J}$) with the property that 
\begin{eqnarray}\label{zero_dot_product}
 \bm{Y}_\perp^i \cdot \bm{Y}_\perp^j =0, ~~ \forall i \neq j, ~~i, j \in \mathcal{J}. 
 \end{eqnarray}
 The cache-configuration vector $\bm{Y}_\perp$ is constructed by first sorting the vector $\bm{v} \equiv \sum_{t=1}^{T} \bm{X}_t$ in non-increasing order. Then, we let the cache configuration $\bm{Y}^j_\perp$ correspond to the set of $C$ files from rank $(j-1)C+1$ to $jC$.   
Clearly, the above construction ensures property \eqref{zero_dot_product}. 
Let the operator $S_{\bm v}(m,n)$ denote the sum of the coordinates running from $m$ to $n$ of the vector $\bm{v}$, sorted in non-increasing order.  With the above construction, since all users make the same file request at each time (\emph{i.e.,} $\bm{X}_t^i=\bm{X}_t, \forall i$), we have  $\forall j\in \mathcal{J}$:
\begin{eqnarray} \label{inelastic_pf1}
  \bm{Y}_\perp^j\cdot \bigg( \sum_{i \in \partial^-(j)}\sum_{t=1}^{T} \bm{X}_t^i\bigg) 
 &=& d \bm{Y}_\perp^j \cdot \bm v \nonumber \\&=& dS_{\bm v}\bigg((j-1)C+1, jC \bigg). \nonumber \\
 \end{eqnarray} 
Finally, the reward accrued by the optimal stationary policy $\pi^*$ may be lower-bounded as:
\begin{eqnarray} \label{inelastic_star}
 G_T^{\pi^*} &\stackrel{(a)}{\geq}& \mathbb{E}\bigg( \sum_{j \in \mathcal{J}} \bm{Y}_\perp^j\cdot \big( \sum_{i \in \partial^-(j)}\sum_{t=1}^{T} \bm{X}_t^i\big) \bigg) \nonumber \\
 &\stackrel{(b)}{=}& \sum_{j \in \mathcal{J}} \mathbb{E}\bigg(\bm{Y}_\perp^j\cdot \big( \sum_{i \in \partial^-(j)}\sum_{t=1}^{T} \bm{X}_t^i\big) \bigg) 	\nonumber \\
 &\stackrel{(c)}{=}& d\sum_{j \in \mathcal{J}} \mathbb{E}\bigg(S_{\bm v}\big( (j-1)C+1, jC\big) \bigg) \nonumber\\
  &=& d\mathbb{E}\bigg(\sum_{j \in \mathcal{J}} S_{\bm v}\big( (j-1)C+1, jC\big) \bigg)\nonumber\\
  &=& d\mathbb{E}\big(S_{\bm{v}}(1, |\mathcal{J}|C)\big) \nonumber\\
  &\stackrel{(d)}{\geq} & \frac{Td}{2} + d\sqrt{\frac{|\mathcal{J}|CT}{2\pi}} - \Theta(\frac{1}{\sqrt{T}}),
\end{eqnarray}
where the inequality (a) follows from Eqn.\ \eqref{obs_2} of observation 2, the equation (b) follows from the linearity of expectation, and the inequality (c) follows from Eqn.\ \eqref{inelastic_pf1}. Finally, we realize that the random variable $S_{\bm{v}}(1, |\mathcal{J}|C)$ is distributed as the total load of the most loaded $|\mathcal{J}|C$ bins when a total of $T$ balls are thrown uniformly at random to $2 |\mathcal{J}|C$ bins. Hence, the inequality (d) follows from an application of Lemma \ref{max_load}. Finally, combining Eqns.\ \eqref{inelastic_lb1} and \eqref{inelastic_star}, we have 
\begin{eqnarray*}
R^\pi_T \geq G^{\pi^*}_T - G^{\pi}_T 
\geq d\sqrt{\frac{|\mathcal{J}|CT}{2\pi}} - \Theta(\frac{1}{\sqrt{T}}).	~~\blacksquare
\end{eqnarray*} 

\bibliographystyle{unsrt}
\bibliography{OCO}
\section{Appendix} \label{appendix}
\subsection{Calculations for the Mean Deviation bound of Theorem \ref{reg_lb_spl}, Eqn.\ \eqref{MD_bound}} \label{MD_bound_proof}
We recall Robbin's form of Stirling's formula \cite{robbins1955remark}, which will be useful in obtaining a non-asymptotic bound for regret. 
\begin{eqnarray}\label{stir}
\sqrt{2 \pi}n^{n+\frac{1}{2}}e^{-n}e^{\frac{1}{12n+1}}	\leq n! \leq \sqrt{2 \pi}n^{n+\frac{1}{2}}e^{-n}e^{\frac{1}{12n}}. 
\end{eqnarray}
 Consider the following two possible cases. \\ 
 \textbf{Case I: $T$ is even}\\
In this case, we lower bound the Mean absolute deviation in Eqn.\ \eqref{DeMoivre} as follows:
\begin{eqnarray*}
	\mathbb{E}|Z-\frac{T}{2}|&=&\frac{2}{2^{T+1}}\big(\lfloor \frac{T}{2} \rfloor +1 \big)\binom{T}{\lfloor \frac{T}{2} \rfloor +1} \\
	&=& \frac{1}{2^T}\frac{T!(\frac{T}{2})}{(\frac{T}{2})!(\frac{T}{2})!}\\
	&\stackrel{(a)}{\geq} & \frac{1}{2^T}\frac{1}{\sqrt{2\pi}}(\frac{T}{2}) \frac{T^{T+\frac{1}{2}}}{(\frac{T}{2})^{T+1}}  e^{\frac{1}{12T+1}-\frac{1}{3T}}\\
	&=& \frac{\sqrt{T}}{\sqrt{2\pi}}e^{-\frac{9T+1}{3T(12T+1)}}\\
	&\stackrel{(b)}{\geq}& \sqrt{\frac{T}{2\pi}}-\frac{9T+1}{3\sqrt{2\pi T}(12T+1)}
\end{eqnarray*}
where the inequality (a) follows from the bound \eqref{stir} and the inequality (b) follows from the fact that $e^{-x} \geq (1-x)$ for all $x \in \mathbb{R}$. Using the fact that $\frac{9T+1}{12T+1} < 1$  for all $T>0$, the mean deviation can be lower bounded as:
\begin{equation} \label{T_even}
\mathbb{E}|Z-\frac{T}{2}| \geq \sqrt{\frac{T}{2\pi}}-\frac{1}{3\sqrt{2\pi T}}
\end{equation}

\textbf{Case II: $T$ is odd}  \\
In this case, the binomial coefficient $\binom{T}{\lfloor \frac{T}{2} \rfloor +1}$ may be lower bounded as follows:
\begin{eqnarray*}
	&&\binom{T}{\lfloor \frac{T}{2} \rfloor +1}\\
	&=& \frac{T!}{(\frac{T+1}{2})!(\frac{T-1}{2})!}\\
	&\geq & \frac{1}{\sqrt{2\pi}} \frac{T^{T+\frac{1}{2}}}{(\frac{T^2-1}{4})^{\frac{T}{2}}}\frac{2}{T+1}e^{-\frac{1}{T}(\frac{1}{6+\frac{6}{T}}+\frac{1}{6-\frac{6}{T}}-\frac{1}{12+\frac{1}{T}})}.
\end{eqnarray*}

Note that 
\[ \frac{T^{T+\frac{1}{2}}}{(\frac{T^2-1}{4})^{\frac{T}{2}}} \geq \frac{T^{T+\frac{1}{2}}}{(\frac{T^2}{4})^{\frac{T}{2}}}= 2^{T} \sqrt{T}. \]
Also, for $T\geq 3$
\[ \frac{1}{6+\frac{6}{T}}+\frac{1}{6-\frac{6}{T}}-\frac{1}{12+\frac{1}{T}} \leq \frac{1}{2}. \]

This gives us the following lower bound
\[ \binom{T}{\lfloor \frac{T}{2} \rfloor +1} \geq \frac{1}{\sqrt{2\pi}} \frac{2^{T+1} \sqrt{T}}{T+1}(1 -\frac{1}{2T}).\]

For $T=1$, the inequality holds good by direct inspection. Hence, the mean deviation from Eqn.\ \eqref{DeMoivre} is lower bounded as
\begin{eqnarray} \label{T_odd}
	\mathbb{E}|Z-\frac{T}{2}| &\geq& \frac{1}{2^T}\big(\frac{T+1}{2}\big)\frac{1}{\sqrt{2\pi}} \frac{2^{T+1} \sqrt{T}}{T+1}(1 -\frac{1}{2T})\nonumber \\
	 &\geq& \sqrt{\frac{T}{2\pi}}- \frac{1}{2\sqrt{2\pi T}}.
\end{eqnarray}

Finally, combining equations \ \eqref{T_even} and \eqref{T_odd} together, we have the following lower bound for all $T\geq 1$

\[\mathbb{E}|Z-\frac{T}{2}|\geq  \sqrt{\frac{T}{2\pi}}-\frac{1}{2\sqrt{2\pi T}},\]
which verifies Eqn.\ \eqref{MD_bound}.

\subsection{Proof of Theorem \ref{elastic_uncoded}}\label{elastic_uncoded_pf}
Let $R_T^{\textrm{single}}$ denote the regret of a single cache under the FTPL caching policy. Due to linearity of the reward functions for elastic contents, the regret for bipartite network caching may be simply upper bounded by summing the regret for each of the $|\mathcal{J}|$ constituent single caches, \emph{i.e.,}
\begin{eqnarray}\label{comb_reg}
 R_T \leq |\mathcal{J}|R_T^{\textrm{single}}.
\end{eqnarray}
 However, note that we can not use the regret upper bound from Theorem \ref{uc_achievability} to upper bound the regret of the individual caches, as, unlike the setting of Theorem \ref{uc_achievability}, each individual caches may receive up to $d$ requests from its neighboring users per slot.  Hence, we need to modify the proof of Theorem \ref{uc_achievability} to take up to $d$ requests into account. Following the same line of argument as in the proof of Theorem \ref{uc_achievability_proof}, the inequality \eqref{quad_form1} may be replaced by the following 
 \begin{eqnarray} \label{quad_form2}
 \langle \bm{x}_t, \nabla^2\Phi_\eta(\tilde{\bm{x}_t})\bm{x}_t)\rangle \leq d^2\max_{i,j,\bm{x}} \big(|\nabla^2\Phi_\eta(\bm{x})\big|)_{ij}.
\end{eqnarray}
Eqn.\ \eqref{quad_form2} follows from the fact that since the number of non-zero terms in $\bm{x}_t$ is at most $d$ (\emph{i.e.,} $||\bm{x}_t||_0 \leq d, ||\bm{x}_t ||_1 \leq d$). Finally, proceeding similarly as in Eqn.\ \eqref{hess_bd}, we conclude that 
\[ \big(|\nabla^2\Phi_\eta(\bm{x})\big|)_{ij} \leq \frac{1}{\eta}\sqrt\frac{2}{\pi}, ~~\forall \bm{x}, i, j.\]
Plugging in the above in Eqn.\ \eqref{main_bd} results in the following regret bound for the \textsf{FTPL} caching policy:
\begin{eqnarray}\label{ind_reg}
 \mathbb{E}(R_T^{\textrm{single}}) \leq C\eta \sqrt{2 \log N}+ d^2\frac{T}{\eta\sqrt{2\pi}}.
\end{eqnarray}
Choosing $\eta = \frac{d}{(4 \pi \log N)^{1/4}}\sqrt{\frac{T}{C}},$ and combining Eqns \eqref{comb_reg} and \eqref{ind_reg}, we have the following regret upper bound for the \textsf{FTPL} policy for caching elastic contents in a $d$-right regular bipartite caching network:
\begin{eqnarray*}
	\mathbb{E}_{\{\bm{\gamma}_t\}_t}(R_T) \leq  1.51(\log N)^{1/4}d|\mathcal{J}|\sqrt{CT}.
\end{eqnarray*}
\subsection{Additional Plots}
In this Section, we include two additional experimental results above. 
\begin{figure}
\centering
\begin{overpic}[width=0.4\textwidth]{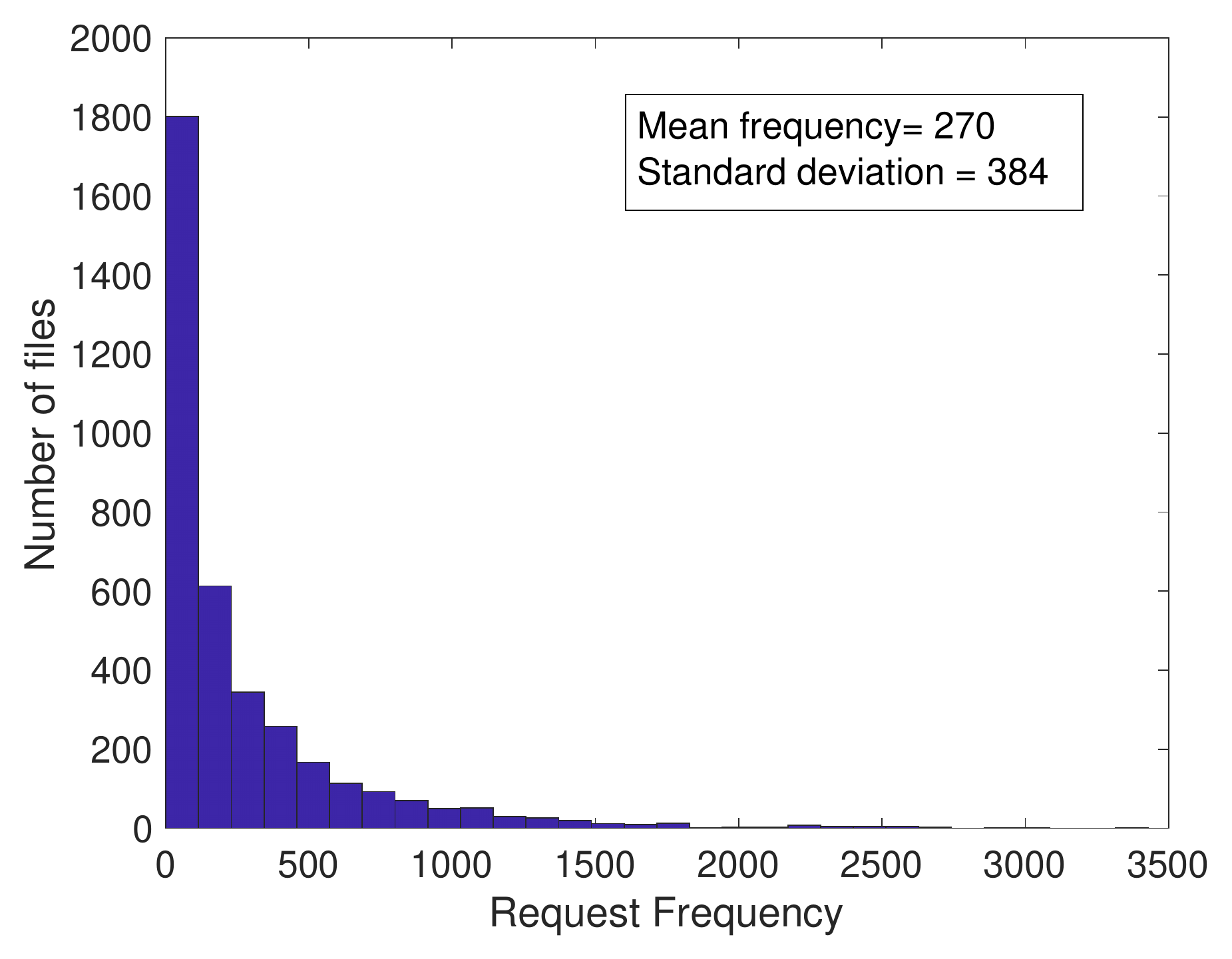}
\end{overpic}
\caption{\small {Histogram of the requests of the MovieLens 1M dataset}}
\label{file_request_histogram}
\end{figure}

\begin{figure}[H]
\centering
\begin{overpic}[width=0.35\textwidth]{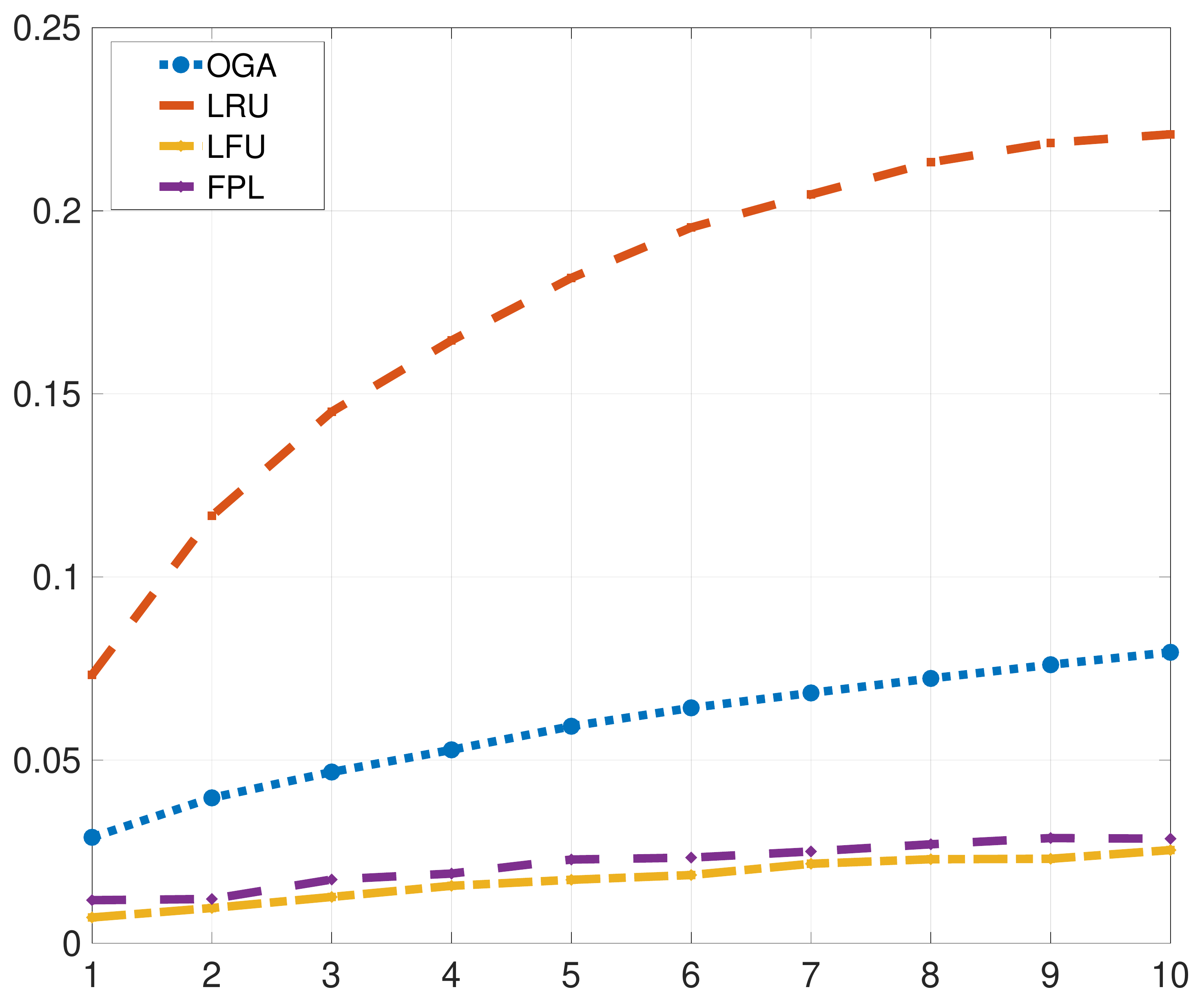}
\put(-10,43){\footnotesize{$\frac{R_T}{T}\bigg|_{T=5\times 10^4}$}}
\put(43,-3){\footnotesize{$C/N$ (in percentage)}}
\end{overpic}
\caption{\small {Variation of the regrets as a function of the cache capacity in the single cache setting for different caching policies}}
\label{C_regret_plot}
\end{figure}

\end{document}